\documentclass[11pt,letter,twocolumn]{IEEEtran}

\usepackage{setspace}
\usepackage{amsmath}
\usepackage{amssymb}
\usepackage{graphicx} 
\usepackage{verbatim}   
\usepackage{color}     
\usepackage{subfigure}  
\usepackage{stmaryrd}
\usepackage{bbm}
\usepackage{url}

\newtheorem{theorem}{Theorem}

\newtheorem{lemma}{Lemma}
\newtheorem{proposition}{Proposition}

\newtheorem{remark}{Remark}

\def\EE{{\mathbb E}}
\def\PP{{\mathbb P}}
\def\RR{{\mathbb R}}

\newcommand{\separator}{
  \begin{center}
    \rule{\columnwidth}{0.3mm}
  \end{center}
}
\newenvironment{separation}{ \vspace{-0.3cm}  \separator  \vspace{-0.2cm}}
{  \vspace{-0.4cm}  \separator  \vspace{-0.1cm}}

\newcommand{\bp}{\noindent{\bf Proof.}\ }
\newcommand{\ep}{\hfill $\Box$}

\newcommand{\BEAS}{\begin{eqnarray*}}
\newcommand{\EEAS}{\end{eqnarray*}}
\newcommand{\BEA}{\begin{eqnarray}}
\newcommand{\EEA}{\end{eqnarray}}
\newcommand{\BEQ}{\begin{equation}}
\newcommand{\EEQ}{\end{equation}}
\newcommand{\BIT}{\begin{itemize}}
\newcommand{\EIT}{\end{itemize}}
\newcommand{\BNUM}{\begin{enumerate}}
\newcommand{\ENUM}{\end{enumerate}}

\newcommand{\eq}[1]{ \begin{equation} #1 \end{equation}}
\newcommand{\eqs}[1]{ \begin{equation*} #1 \end{equation*}}

\newcommand{\als}[1]{ \begin{align*} #1 \end{align*}}
\newcommand{\sk}{ \nonumber\\}
\newcommand{\indic}{ { \bf 1} }
\newcommand{\ouralgo}{CRS-T }
\newcommand{\ouralg}{CRS-T}
 
\newcommand{\ceil}[1]{  \lceil #1 \rceil}

\begin{document}

\title{Dynamic Rate and Channel Selection \\in Cognitive Radio Systems}

\author{R. Combes and A. Proutiere\\ KTH, The Royal Institute of Technology}
\maketitle

\begin{abstract}
In this paper, we investigate dynamic channel and rate selection in cognitive radio systems which exploit a large number of channels free from primary users. In such systems, transmitters may rapidly change the selected (channel, rate) pair to opportunistically learn and track the pair offering the highest throughput. We formulate the problem of sequential channel and rate selection as an online optimization problem, and show its equivalence to a {\it structured} Multi-Armed Bandit problem. The structure stems from inherent properties of the achieved throughput as a function of the selected channel and rate. We derive fundamental performance limits satisfied by {\it any} channel and rate adaptation algorithm, and propose algorithms that achieve (or approach) these limits. In turn, the proposed algorithms optimally exploit the inherent structure of the throughput. We illustrate the efficiency of our algorithms using both test-bed and simulation experiments, in both stationary and non-stationary radio environments. In stationary environments, the packet successful transmission probabilities at the various channel and rate pairs do not evolve over time, whereas in non-stationary environments, they may evolve. In practical scenarios, the proposed algorithms are able to track the best channel and rate quite accurately without the need of any explicit measurement and feedback of the quality of the various channels.
\end{abstract}


\section{Introduction}

In cognitive radio systems, radio devices may access a potentially large number of frequency bands or channels. An example of such systems are those exploiting "white space" spectrum, the unused part of the TV/UHF spectrum (unallocated or not used locally). The FCC 2008 ruling allowed unlicensed devices to use parts of this spectrum, provided that devices can detect primary users (TV transmitters and wireless microphones). As a part of the 2010 ruling \cite{fcc}, FCC mandates the use of a geolocation database to identify which frequencies are free from primary users. By querying the geolocation database, we are guaranteed to obtain a set of channels free from primary transmitters and we avoid the difficult problem of sensing primary users. 

We consider systems exploiting channels known to be free from primary users. For the transmission of each packet, transmitters can select a coding rate from a finite predefined set (as in 802.11 systems for example) and a channel from the set of available channels. The outcome of a packet transmission is random, and the  probabilities of successfully transmitting a packet using the various (channel, rate) pairs are a priori unknown at the transmitter; they need to be learnt based on trial and error. These probabilities can vary significantly and randomly over time and across channels; they also strongly depend on the chosen coding rate. As a consequence, tracking the best (channel, coding rate) pair for transmission may greatly improve the system performance. In this paper, we aim at designing sequential channel and coding rate selection schemes that efficiently track the best available channel and the corresponding coding rate.

As shown in previous works, see e.g. \cite{camp2008,radunovic2011}, RSSI (Receive Signal Strength Indicator) is a poor predictor of channel quality, and hence of the packet successful transmission probabilities. In OFDM systems for example, this stems from the fact that RSSI does not report the individual signal strength experienced on the various sub-carriers. In order to accurately estimate the quality of a wide-band channel, more sophisticated techniques with specific hardware are needed \cite{sen2010,halperin2010}. But these techniques are not typically supported in current commercial radio hardware. Instead, we need to infer the quality of each channel at each transmission rate through probing. Here we consider 802.11-like systems, where the only feedback sent from the receiver to the transmitter is whether a data packet has been successfully received or not. Hence by probing, we mean that several real data packets have to be sent on each channel and at each rate to construct a reliable estimate of the channel quality. In the design of channel and rate selection schemes, we then face a classical exploration vs. exploitation trade-off problem. We need to exploit the (channel, rate) pair that has offered the best throughput so far, whilst constantly exploring other pairs in case one of them is actually optimal. 

We rigorously formulate the design of the optimal sequential (channel, rate) selection algorithms as an online stochastic optimization problem. In this problem, the objective is to maximize the number of packets successfully sent over a finite time horizon. We show that this problem reduces to a Multi-Armed Bandit (MAB) problem \cite{lai1985}. In MAB problems, a decision maker sequentially selects an action (also called an ``arm''), and observes the corresponding reward. Rewards of a given arm are random variables with unknown distribution. The objective is to design sequential action selection strategies that maximize the expected reward over a given time horizon. These strategies have to achieve an optimal trade-off between exploitation (actions that have provided high rewards so far have to be selected) and exploration (sub-optimal actions have to be chosen so as to learn their average rewards). For our (channel, rate) selection problem, the various arms correspond to the decisions available at the transmitter to send packets, i.e., an arm corresponds to a channel and a coding rate. When a (channel, rate) pair is selected for a packet transmission, the reward is equal to $1$ if the transmission is successful, and equal to $0$ otherwise. The average successful packet transmission probabilities at the various (channel, rate) pairs are unknown, and have to be learnt.

The MAB problem corresponding to the design of channel and rate selection mechanisms is referred to as a {\it structured} MAB problem. It differs from classical MAB problems. (i) First, the rewards associated with the various rates on a given channel are stochastically correlated, i.e., the outcomes of transmissions at different rates are not independent: for example, if a transmission at a high rate is successful, it would be also successful at lower rates. (ii) Then, the average throughputs achieved at various rates exhibit natural structural properties. For a given channel, the throughput is an unimodal function of the selected rate. (iii) In addition, most often, on all channels, the packet successful transmission probabilities are close to $1$ at low rates, and abruptly decrease to $0$ as the rate increases. This additional structure, referred to as graphical unimodality, allows us to predict the outcomes of transmissions on various channels. As we demonstrate, correlations and (graphical) unimodality are instrumental in the design of channel and rate selection mechanisms, and can be exploited to learn and track the best (channel, rate) pair quickly and efficiently. Finally, note that most MAB problems consider stationary environments, which, for our problem, means that the successful packet transmission probabilities for the different (channel, rate) pairs do not vary over time. In practice, the transmitter faces a non-stationary environment as these probabilities could evolve over time. We consider both stationary and non-stationary radio environments.

In the case of stationary environments, we derive an upper bound of the expected reward that can be achieved in structured MAB problems. This provides a fundamental performance limit that {\it any} (channel, rate) selection algorithm cannot exceed. This limit quantifies the inevitable performance loss due to the need to explore sub-optimal (channel, rate) pairs. It also indicates the performance gains that can be achieved by devising schemes that optimally exploit the correlations and the structural properties of the MAB problem. We present sequential (channel, rate) selection algorithms that optimally exploit the structural properties of the problem: for our algorithms, we prove that the performance loss due to the need to explore sub-optimal (channel, rate) pairs does not depend on the number of available rates. We also extend our algorithms to non-stationary radio environments. Finally, we evaluate the performance of the proposed algorithms using an office white-space testbed operating in the 500MHz-600MHz band, and simulation experiments.

To our knowledge, the problem of sequential channel and rate selection has only been investigated in \cite{radunovic2011}, where heuristic algorithms have been developed. In contrast, we formulate and solve this problem rigorously, i.e., we provide fundamental upper performance bounds satisfied under any channel and rate selection algorithm and design optimal algorithms that match these bounds.

{\it Contributions.}
\begin{itemize}
\item We formulate the design of (channel, rate) selection algorithms as an online optimization problem, and establish its equivalence to a structured MAB problem.
\item We derive a performance upper bound satisfied by any (channel, rate) selection algorithm, depending on the assumptions made on the structure of the problem -- three scenarios with increasing structure are considered: 1. no structural assumption is made; 2. the throughput on each channel is a unimodal function of the rate; 3. the throughput is a graphically unimodal function of the channel and rate. We also quantify the performance gains that one may achieve by exploiting the structural properties of the problem.
\item We propose three (channel, rate) selection algorithms, one for each of the above scenarios, and analyze their performance in stationary radio environments. We prove that our algorithms optimally exploit the structural properties of the throughputs. 
\item We briefly discuss the extensions of our algorithms to non-stationary radio environments.
\item Finally, we evaluate the performance of our algorithms using simulation experiments. To this aim, we use artificially generated traces, as well as traces extracted from a white space test-bed operating in the UHF radio spectrum .  
\end{itemize}

{\it Paper organization.} The next section is devoted to the related work. In Sections~\ref{sec:prelim} and \ref{sec:mab}, we formulate the problem of sequential selection of channel and rate selection as a {\it structured} bandit problem. Section~\ref{sec:low} presents fundamental upper performance bounds for this problem. In Section~\ref{sec:algo}, asymptotically optimal algorithms are proposed. Section~\ref{sec:nonstat} deals with non-stationary radio environments. Section \ref{sec:num} presents numerical experiments to evaluate the performance of our algorithms.

\section{Related work}

First observe that the joint channel and rate selection problem is considerably more difficult than detecting channels with no primary users as considered in a lot of recent works, see e.g. \cite{kim2008, motamedi2008, liu2010, lai2008medium, lai2011, zhao2007, ahmad2009}. In some of these papers, a MAB framework has been used to design primary user detection algorithms. The presence or the absence of primary users just means that a channel is either good or bad. When selecting both channel and rate, the dimension of the problem becomes larger, and there are multiple and numerous possible channel states. Primary users are not considered in our work, as we assume that transmitters can use a geolocation database to get a list of channels free from primary users \cite{fcc}. 

It should also be observed that most of the work on dynamic spectrum access considers stationary radio environments. In \cite{liu2010, lai2008medium} for example, the authors use classical stochastic control techniques (Markov Decision Processes) to sequentially select a channel for transmission. The underlying assumption is that the environment is stationary, i.e., the packet successful transmission probabilities do not evolve over time. In this paper, both stationary and non-stationary radio environments are explored. Test-bed experiments actually suggest that the environment is non-stationary in practice, even in networks where nodes do not move such as indoor offices,  see e.g. \cite{radunovic2011}.

Our problem resembles the rate adaptation problem in 802.11 systems, see e.g.  \cite{bicket2005, wong2006, deek2013}. But again, our problem has one additional dimension (a channel has to be selected): in turn, the number of available decisions at the transmitter is much larger than in 802.11 systems where only the rate has to be chosen. Rate adaptation algorithms are not applicable when the channel can also be selected for each packet transmission. This is due to the fact that the transmitter does not continuously monitor the same channel (as in 802.11 systems), and has to switch channels often to discover the best (channel, rate) pair as rapidly as possible.

There is an abundant literature on MAB problems, and engineers have applied these problems to dynamic spectrum access \cite{motamedi2008, lai2008medium}, \cite{lai2011}, \cite{tekin2011online}. Most existing theoretical results, see \cite{bubek} for a recent survey, are concerned with {\it unstructured} MAB problems, i.e., problems where the average reward associated with the various decisions are not related. For this kind of problems, Lai and Robbins \cite{lai1985} derived an asymptotic lower bound on regret and also designed optimal sequential decision algorithms. When the average rewards are structured (as this is the case for our problem), the design of optimal decision algorithms is more challenging, see e.g. \cite{bubek}. Non-stationary environments have not been extensively studied in the bandit literature: Most often unstructured MAB only are analyzed, see \cite{kocsis2006,yu2009,garivier2011}.  

To our knowledge, the only work dealing with joint (channel, rate) selection is \cite{radunovic2011}. However there, the structural properties of the corresponding MAB problem had not been identified, and the authors only proposed algorithms based on heuristics. This contrasts with the present work: we rigorously determine fundamental limits satisfied by any (channel, rate) adaptation algorithm, and propose algorithms approaching these limits.

\section{Models} \label{sec:prelim}

We consider a single link (a transmitter-receiver pair). At
time 0, the link becomes active and the transmitter starts sending packets to the receiver. For each packet, the transmitter selects a channel from a finite set ${\cal C}=\{1,\ldots,C\}$, and a coding and modulation scheme from a finite set ${\cal K}=\{1,\ldots,K\}$. The transmission rate when using coding and modulation scheme $k$ is denoted by $r_k$ and we define the set of rates ${\cal R} = \{r_k ,k \in {\cal K}\}$. ${\cal R}$ is ordered, i.e., $r_1<r_2<\ldots<r_K$. After a packet is sent, the transmitter is informed of whether the transmission has been
successful. Based on the observed past transmission successes and
failures at the various channels and rates, the transmitter has to select a channel and rate pair for the next packet transmission. Let $\Pi$ denote the set of all possible sequential (channel, rate) selection schemes. Packets are assumed to be of equal size, and without loss of generality,  for any $k$, the duration of a packet transmission at rate $r_k$ is $1/r_k$.

\subsection{Channel models}

For the $i$-th packet transmission on channel $c$ at rate $r_k$, a binary random variable $X_{ck}(i)$ represents the success ($X_{ck}(i)=1$) or failure ($X_{ck}(i)=0$) of the transmission. $\mathbb{E}[X_{ck}(i)]$ refers to as the packet successful transmission probability on channel $c$ at rate $r_k$ (i.e., it is the packet reception rate). We consider both stationary and non-stationary radio environments. In stationary environments, the success transmission probabilities on the various channels and at different rates do not evolve over time. This arises when the system considered is static (in particular, the transmitter and receiver do not move). In non-stationary environments, success transmission probabilities can evolve over time. Unless otherwise specified, we consider stationary radio environments. Non-stationary environments are treated in Section \ref{sec:nonstat}.  

We assume that $X_{ck}(i)$, $i=1,2,\ldots$, are independent and identically distributed, and we denote by $\theta_{ck}$ the success transmission probability on channel $c$ at rate $r_k$: $\theta_{ck}=\mathbb{E}[X_{ck}(i)]$. We verified that the i.i.d. assumption holds in our test-bed and simulation framework. Denote by $(c^\star,k^{\star})$ the optimal (channel, rate) pair, i.e., $(c^\star,k^{\star}) \in \arg\max_{c,k} r_k\theta_{ck}$. To simplify the exposition and the notation, we assume that the optimal (channel, rate) pair is unique, i.e., $r_{k^{\star}}\theta_{c^\star k^{\star}}>r_k\theta_{ck}$, for all $(c,k)\neq (c^\star,k^\star)$. Our analysis can be extended in an easy way to scenarios where the optimal channel and rate pair is not unique. Note however that scenarios where different channel and rate pairs yield exactly the same throughput should happen very rarely in practice. We further introduce, for any channel $c$, the optimal rate $r_{k_c^\star}$, i.e., $(c,k_c^{\star}) \in \arg\max_{k} r_k\theta_{ck}$. Again for simplicity, we assume that on any channel, the optimal rate is unique: $r_{k_c^{\star}}\theta_{c k_c^{\star}}>r_k\theta_{ck}$, for all $k\neq k_c^\star$. The throughput achieved using (channel, rate) pair $(c,k)$ is denoted by $\mu_{c k} = r_{k} \theta_{c k}$. The maximum throughput on channel $c$ is $\mu^\star_c = \mu_{c k^\star_c }$, and the throughput achieved using the optimal (channel, rate) pair is $\mu^\star = \mu^\star_{c^\star} = \mu_{c^\star k^\star}$.

Although we do not account for the presence of primary users in this work, we could actually model scenarios where on each channel $c$, primary users occupy the channel with some fixed probability $\zeta_c$, and in an i.i.d. manner across time. Indeed in such scenarios, we just need to replace $\theta_{ck}$ by $(1-\zeta_c)\theta_{ck}$. If the primary users occupy channels for long periods of time (not in an i.i.d. manner), the analysis would be significantly more challenging. This kind of situations is investigated in \cite{liu2010} for example. 

\subsection{Structural properties}

The successful transmission probabilities $\theta=(\theta_{ck}, c\in {\cal C}, k\in {\cal K})$ are initially unknown at the transmitter, and have to be learnt. When the number of (channel, rate) pairs grows large, learning the best pair for transmission then becomes really challenging. Fortunately, the outcomes of transmissions using the various (channel, rate) pairs exhibit structural properties that can be exploited to speed up the learning process. To emphasize the importance of exploiting the structural properties, we consider three scenarios with increasing structure.

\subsubsection{Scenario 1 -- No structure}
If no structural assumptions are made regarding the successful transmission probabilities, then $\theta\in [0,1]^{C\times K}$. In such scenarios, we will show that the performance loss due to the need to explore sub-optimal (channel, rate) pairs scales linearly with the number of channels and rates.

\subsubsection{Scenario 2 -- Unimodality}

First observe that the successes and failures of transmissions on a given channel at various rates are statistically correlated. Indeed, if a transmission is successful at a high rate, it has to be successful at a lower rate. Similarly, if a low-rate transmission fails, then transmitting at a higher rate would also fail. Formally this means that for any channel $c$, $\theta_c=(\theta_{c1},\ldots,\theta_{cK})\in {\cal T}$, where ${\cal T}=\{\eta\in [0,1]^K: \eta_1\ge \ldots\ge \eta_K\}$. Then, in practice, it has been observed (and this is confirmed in our numerical experiments) that the throughput achieved on a given channel is a unimodal function of the transmission rate, see e.g. \cite{halperin2010, deek2013}. In other words, for any channel $c$, $\theta_c\in {\cal U}$, where ${\cal U}=\{\eta\in [0,1]^K:\exists k_1, r_1\eta_1<\ldots <r_{k_1}\eta_{k_1}, r_{k_1}\eta_{k_1}> r_{k_1+1}\eta_{k_1+1}>\ldots>r_K\eta_K\}$. In summary in Scenario 2, for any channel $c$, $\theta_c\in {\cal T}\cap {\cal U}$. 

\subsubsection{Scenario 3 -- Graphical unimodality}\label{not1}

We further observe (see Section \ref{sec:num}) that on a given channel, the throughput first grows linearly with the rate (the successful transmission probability is close to 1), and then abruptly decreases to 0. This observation has been made in earlier work, see \cite{bicket2005} (the author refers to this scenario as the {\it steep throughput} scenario), \cite{halperin2010}. This knowledge can be exploited to build a relationship between the throughputs achieved on various channels. Indeed, for example, the throughputs observed on two different channels are roughly identical in their growth phase (when the rates are low and the success probabilities are close to 1). To exploit this observation, we remark that if it holds, the throughput is a {\it graphically unimodal} function of the (channel, rate) pair as defined below.

We first construct a directed graph $G=(V,E)$ whose vertices correspond to the (channel, rate) pairs. When $(d,d')\in E$, we say that the decision $d'$ is a neighbor of decision $d$, and we define ${\cal N}(d)=\{d'\in V: (d,d')\in E\}$ as the set of neighbors of $d$. The throughput or average reward of decision $d=(c,k)$ is denoted by $\mu_d=r_k\theta_{ck}$. Graphical unimodality expresses the fact that when the optimal decision is $d^\star=(c^\star,k^\star)$, then for any $d\in V$, there exists a path in $G$ from $d$ to $d^\star$ along which the expected reward is strictly increasing. In other words there is no {\it local} maximum in terms of expected reward except at $d^\star$. The notion of locality is defined through that of neighborhood ${\cal N}(d), d\in V$. Formally, $\theta\in {\cal U}_G$, where ${\cal U}_G$ is the set of successful transmission probabilities $\theta\in [0,1]^{C\times K}$ such that, if $d^\star=(c^\star,k^\star)\in \arg\max_{(c,k)} r_k\theta_{ck}$, for any $d=(c,k)\in V$, there exists a path $(d_0=d,d_1,\ldots,d_p=d^\star)$ in $G$ such that for any $i=1,\ldots,p$, $\mu_{d_i} > \mu_{d_{i-1}}$. 

\begin{figure}[htb]
\begin{center}
\includegraphics[width=4.5cm]{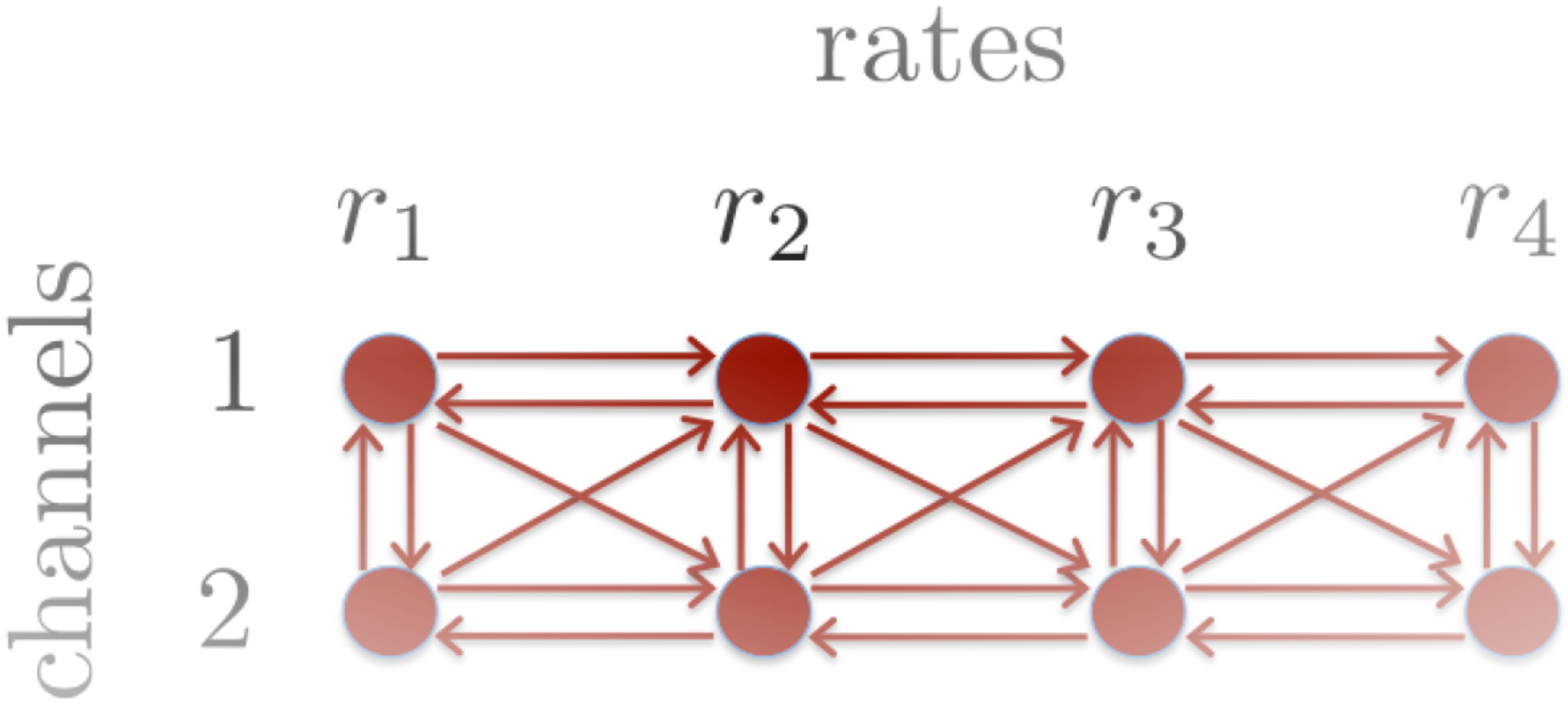} 
\end{center}
\vspace{-4mm}
\caption{Example of a graph providing unimodality of the throughput.}
\label{fig1}
\end{figure}

Let us now complete the construction of graph $G$. The set of edges $E$ is: $((c,k),(c,k-1))$, $((c,k), (c,k+1))$ and $((c,k),(c',k))$, $((c,k),(c',k+1))$ for all (channel, rate) pair $(c,k)$, and all $c'$. An example of such a graph $G$ is presented in Figure \ref{fig1} -- for 2 channels and 4 rates. When the above observation made on $\theta$ holds (steep scenario as defined in \cite{bicket2005}), it is easy to check that the throughput is a graphically unimodal function (w.r.t. graph $G$) of the channel and rate. In all practical cases, beyond the steep throughput scenario, we have actually observed that the graph $G$ as constructed above had enough edges to guarantee the graphical unimodality of the throughput, see Section \ref{sec:num}.

In summary, in Scenario 3, we assume that $\theta\in {\cal T}^C\cap {\cal U}_G$. Note that there is more structure in Scenario 3 than in Scenario 2: if $\theta\in  {\cal T}^C\cap {\cal U}_G$, then for any $c$, $\theta_c\in {\cal T}\cap {\cal U}$.

\begin{remark} Note that the sharp decrease of the throughput when the rate increases may not hold in some scenarios as observed in several papers. The sharp transition of the reception rate motivates the use of graphical unimodality, but the latter is more general, and may hold even in absence of this sharp transition. Actually, we are free to design the graph $G$, and adapt its topology depending on the specificity of the radio environment so as to get graphical unimodality.
\end{remark}

\begin{remark} The notion of graphical unimodality is generic. Our approach consists in constructing a minimal graph $G$ combining structural properties satisfied by the throughput as a function of the (channel, rate) pair, and results from experiments run off-line, so that the throughput is graphically unimodal w.r.t. $G$. It should be observed that in MIMO systems (e.g. as in 802.11n and subsequent standards), the throughput is no longer a unimodal function of the rate (due to the different available MIMO modes, with single or multiple streams). The proposed framework can be adapted to account for the various MIMO modes. To this aim, the set of decisions would correspond to the set of all possible (channel, MIMO mode, rate) triplets, and the graph $G$ would be constructed so that the throughput is a graphically unimodal function of these triplets.
\end{remark}

\section{Objectives and Multi-Armed Bandits}\label{sec:mab}

Our goal is to devise a sequential (channel, rate) selection scheme that maximizes the number of packets successfully transmitted over a finite time horizon. Such a design can be formulated as an online stochastic optimization problem. The choice of the time horizon, denoted by $T$, is not really important as long as during time interval $T$, a large number of packets can be sent -- so that inferring the success transmission probabilities efficiently is possible. 

Consider a rate adaption scheme $\pi\in \Pi$ that selects (channel, rate) pair $(c^{\pi}(t),k^{\pi}(t))$ for the $t$-th packet transmission. The number of packets $\gamma^\pi(T)$ that have been successfully sent under algorithm $\pi$ up to time $T$ is: $\gamma^\pi(T) = \sum_{c,k} \sum_{i=1}^{s_{ck}^\pi(T)}X_{ck}(i)$, where $s_{ck}^\pi(T)$ is the number of transmission attempts on channel $c$ at rate
$r_k$ before time $T$. The $s_{ck}(T)$'s are random variables (since the
rates selected under $\pi$ depend on the past random successes and
failures), and satisfy the following constraint:
$$
\sum_{c,k}s_{ck}^\pi(T)\times {1\over r_k} \le T.
$$
Wald's lemma implies that the expected number of packets
successfully sent up to time $T$ is: $\mathbb{E}[\gamma^\pi(T)]=\sum_{c,k}
\mathbb{E}[s_{c,k}^{\pi}(T)] \theta_{ck}.$ Thus, our objective is to design
an algorithm solving the following online stochastic optimization
problem:
\begin{eqnarray}\label{pb:1}
\max_{\pi\in \Pi} & \sum_{c,k} \mathbb{E}[s_{ck}^{\pi}(T)] \theta_{ck},\\
\hbox{s.t. } & \sum_{c,k} s_{ck}^{\pi}(T)\times {1\over r_k} \le T,\quad\forall c,k, s_{ck}^\pi(T)\in \mathbb{N}.\nonumber
\end{eqnarray}

\subsection{An equivalent Multi-Armed Bandit (MAB) problem}

Next we show that the above online stochastic optimization problem is equivalent to a Multi-Armed Bandit (MAB) problem.

\subsubsection{An alternative system}

Without loss of generality, we assume that time can be divided into slots whose durations are such that for any $k$, the time it takes to transmit one packet at rate $r_k$ corresponds to an integer number of slots. Under this convention, the optimization problem (\ref{pb:1}) can be written as:
\begin{eqnarray}\label{pb:1bis}
\max_{\pi\in \Pi} & \sum_{c,k} \mathbb{E}[t_{ck}^{\pi}(T)] r_k\theta_{ck},\\
\hbox{s.t. } & \sum_{c,k} t_{ck}^{\pi}(T)\le T,\nonumber\\
& \forall c, k, t_{ck}^\pi(T)\in {1\over r_k}\mathbb{N}=\{ {u\over r_k}, u\in \mathbb{N}\},\nonumber
\end{eqnarray}
where $t_{ck}^\pi(T)=s_{ck}^\pi(T)/r_k$ represents the amount of time (in slots) that the transmitter spends, before $T$, on sending packets on channel $c$ at rate $r_k$. The constraint $t_{ck}(T)\in {1\over r_k}\mathbb{N}$ indicates that when a rate is selected, this rate selection remains the same for the next $1/r_k$ slots. By relaxing this constraint, we obtain an optimization problem corresponding to a MAB problem. Indeed, consider now an alternative system where rate selection is made {\it every} slot. If at any given slot, (channel, rate) pair $(c,k)$ is selected for the $i$-th times, then if $X_{ck}(i)=1$, the transmitter successfully sends $r_k$ bits in this slot, and if $X_{ck}(i)=0$, then no bit are received. A (channel, rate) selection algorithm then decides in each slot which (channel, rate) pair to use. There is a natural mapping between rate selection algorithms in the original system and in the alternative system: let $\pi\in \Pi$, if for the $t$-th packet transmission, rate $r_k$ is selected under $\pi$ in the original system, then $\pi$ selects the same rate $r_k$ in the $t$-th slot. 



For the alternative system, the objective is to design $\pi\in \Pi$ solving the following optimization problem, which can be interpreted as a relaxation of (\ref{pb:1bis}). 
\begin{eqnarray}\label{pb:2}
\max_{\pi\in \Pi} & \sum_{c,k} \mathbb{E}[t_{ck}^{\pi}(T)] r_k\theta_{ck},\\
\hbox{s.t. } & \sum_{c,k} t_{ck}^{\pi}(T)\le T\nonumber,\\
& \forall c, k, t_{ck}^{\pi}(T)\in \mathbb{N}.\nonumber
\end{eqnarray}
The above optimization problem corresponds to a MAB problem, where in each slot a decision is taken (i.e., a channel and a rate are selected), and where when $(c,k)$ is chosen, the obtained reward is $r_k$ with probability $\theta_{ck}$ and 0 with probability $1-\theta_{ck}$. 

\subsubsection{Regrets}

We quantify the performance of an algorithm $\pi\in\Pi$ in both original and alternative systems through the notion of {\it regret}. The regret up to slot $T$ compares the performance of $\pi$ to that achieved by an algorithm always selecting the best (channel, rate) pair. If the parameter $\theta=(\theta_{ck},c,k)$ was known, then in both systems, it would be optimal to always select (channel, rate) pair $(c^\star, k^\star)$. The regret of algorithm $\pi$ up to time slot $T$ in the original system is then defined by:
$$
R_1^{\pi}(T) = \theta_{c^\star k^\star}\lfloor r_{k^\star}T\rfloor -\sum_{c,k}\theta_{ck}\mathbb{E}[s_{ck}^{\pi}(T)],
$$
where $\lfloor x\rfloor$ denotes the largest integer smaller than $x$. 

The regret of algorithm $\pi$ up to time slot $T$ in the alternative system is similarly defined by:
$$
R^{\pi}(T) = \theta_{c^\star k^\star}r_{k^\star}T -\sum_{c,k}\theta_{ck}r_k \mathbb{E}[t_{ck}^{\pi}(T)].
$$

\subsubsection{Asymptotic equivalence}

In the next section, we show that an asymptotic lower bound for the regret $R^\pi(T)$ is of the form $c(\theta)\log(T)$ where $c(\theta)$ is a strictly positive constant that we can explicitly characterize. It means that for all $\pi\in \Pi$, $\lim\inf_{T\to\infty}R^\pi(T)/\log(T)\ge c(\theta)$. It can be also shown that there exists an algorithm $\pi^\star\in \Pi$ that actually achieves this lower bound in the alternative system, in the sense that $\lim\sup_{T\to\infty}R^{\pi^{\star}}(T)/\log(T)\le c(\theta)$. In such a case, we say that $\pi^\star$ is asymptotically optimal. The following proposition states that actually, the same lower bound is valid in the original system, and that any asymptotically optimal algorithm in the alternative system is also asymptotically optimal in the original system. 

\begin{proposition}
Let $\pi\in \Pi$. For any $\beta >0$, we have:
$$
 \lim\inf_{T\to\infty}{R^{\pi}(T)\over \log(T)}\ge \beta \Longrightarrow \lim\inf_{T\to\infty}{R_1^{\pi}(T)\over \log(T)}\ge \beta ,
$$
and
$$
\lim\sup_{T\to\infty}{R^{\pi}(T)\over \log(T)}\le \beta \Longrightarrow \lim\sup_{T\to\infty}{R_1^{\pi}(T)\over \log(T)}\le \beta .
$$
\end{proposition}

\medskip
\bp Let $T>0$. By time $T$, we know that there have been at least $\lfloor T r_1\rfloor$ transmissions, but no more than $\lceil Tr_K\rceil$. Also observe that both regrets $R^{\pi}$ and $R^{\pi}_1$ are increasing functions of time. We deduce that:
$$
R^{\pi}(\lfloor Tr_1\rfloor)\le R_1^{\pi}(T) \le  R^{\pi}(\lceil Tr_K\rceil).
$$
Now 
\begin{align*}
\lim\inf_{T\to\infty}{R_1^{\pi}(T)\over \log(T)} & \ge \lim\inf_{T\to\infty}{R^{\pi}(\lfloor Tr_1\rfloor)\over \log(T)}\\
& = \lim\inf_{T\to\infty}{R^{\pi}(\lfloor Tr_1\rfloor)\over \log(\lfloor Tr_1\rfloor)} \ge \beta.
\end{align*}
The second statement can be derived similarly.
\ep
\medskip

\subsection{MAB problems}

Instead of trying to solve (\ref{pb:1}), we rather focus on analyzing the MAB problem (\ref{pb:2}). We know that optimal algorithms for (\ref{pb:2}) will also be optimal for the original problem. The nature of the MAB problem (\ref{pb:2}) depends on the structural assumption made on the successful transmission probabilities $\theta$. In absence of such assumption (Scenario 1), we get a classical MAB problem where the rewards provided by all decisions are independent. In Scenarios 2 and 3, we get a {\it structured} MAB problem, as we know a priori that $\theta$ belongs to a structured set, which helps learning the best (channel, rate) pair. Next we summarize the MAB problems obtained in the three different scenarios.

We have a set $\{1,\ldots,C\}\times \{1,\ldots,K\}$ of possible decisions (i.e., (channel,rate) pairs). If decision $(c,k)$ is taken for the $i$-th time, we receive a reward $r_kX_{ck}(i)$. $(X_{ck}(i),i=1,2,...)$ are i.i.d. with Bernoulli distribution with mean $\theta_{ck}$. The objective is to design a decision scheme minimizing the regret $R^\pi(T)$ over all possible algorithms $\pi\in \Pi$. The three MAB problems differ depending on the structural assumptions made on $\theta$.

\medskip
\noindent
{\bf Unstructured MAB $(P_I)$}. No assumption is made on $\theta$: $\theta\in [0,1]^{C\times K}$.
 
\medskip
\noindent
{\bf Structured MAB $(P_U)$}. We assume that $\theta_c\in {\cal T}\cap {\cal U}$ for all channel $c$.

\medskip
\noindent
{\bf Structured MAB $(P_{GU})$}. We assume that $\theta\in {\cal T}^C\cap {\cal U}_G$.

\section{Regret Lower Bounds}\label{sec:low}

In this section, we derive an asymptotic (as $T$ grows large) lower bound of the regret $R^{\pi}(T)$ satisfied by any algorithm $\pi\in \Pi$ in the three MAB bandit problems $(P_I)$, $(P_U)$, and $(P_{GU})$. These lower bounds provide insightful theoretical performance limits satisfied by any (channel, rate) selection scheme. By comparing the lower bounds derived for the three problems, we also quantify the performance gains that can be achieved by smartly exploiting the (a priori) known structure. 

\subsection{Unstructured MAB $(P_I)$}

The regret lower bound for MAB problem $(P_I)$ can be derived using the direct technique used by Lai and Robbins \cite{lai1985}. Note that the only difference between $(P_I)$ and the classical MAB problems \cite{lai1985} lies in the fact that in $(P_I)$, we know that the average reward of decision $(c,k)$ is of the form $r_k\theta_{ck}$ for known $r_k$. The analysis of $(P_I)$ is then similar to that of classical bandit problems. 

We first introduce the notion of uniformly good algorithms. An algorithm $\pi$ is uniformly good, if for all parameters $\theta$, for any $\alpha >0$, we have\footnote{$f(T)=o(g(T))$ means that $\lim_{T\to\infty} f(T)/g(T)=0$.}: $\mathbb{E}[t_{ck}^{\pi}(T)]=o(T^\alpha), \forall (c,k) \neq (c^\star, k^{\star})$, where $t_{ck}^{\pi}(T)$ is the number of times (channel, rate) pair $(c, k)$ has been chosen up to the $T$-th decision, and $(c^\star,k^\star)$ is the optimal channel and rate pair (it depends on $\theta$). Uniformly good algorithms exist as we shall see later on. 

Let $N=\{k:  \mu^\star \le r_k\}$ -- note that $N$ depends on $\theta$. There exists $k_0$ such that $N=\{k_0,\ldots,K\}$, with the convention $k_0=K+1$ if $N=\emptyset$. Note that if $k<k_0$, then for any channel $c$, $r_{k^{\star}}\theta_{ck^{\star}} > r_k$, which means that even if all transmissions at rate $r_k$ on channel $c$ were successful, i.e., $\theta_{ck}=1$, rate $r_k$ would be sub-optimal. Hence, there is no need to select rate $r_k$ to discover this fact, since by only selecting rate $r_{k^{\star}}$ on channel $c^\star$, we get to know whether $r_{k^{\star}}\theta_{c^\star k^{\star}} > r_k\ge r_k\theta_k$. 

Finally, we introduce the Kullback-Leibler (KL) divergence, a well-known measure for dissimilarity between two distributions. When we compare two Bernoulli distributions with respective averages $p$ and $q$, the KL divergence is: $I(p,q) = p \log \frac{p}{q} + (1-p) \log \frac{1-p}{1-q}$. 

\medskip
\begin{theorem}\label{th:lower_indep}
Let $\pi\in \Pi$ be a uniformly good rate selection algorithm for MAB problem ($P_I$). We have: $\liminf_{T\to\infty} {R^\pi(T)\over \log(T)} \ge c_I({\theta})$,
where 
\begin{align*}
c_I({\theta}) = & \sum_{k=k_0: k\neq k^\star}^K {\mu^\star - r_k\theta_{c^\star k}\over I(\theta_{c^\star k}, \frac{\mu^\star}{r_k})}+\sum_{c\neq c^\star}\sum_{k=k_0}^K {\mu^\star - r_k\theta_{ck}\over I(\theta_{ck}, \frac{\mu^\star}{r_k})}.
\end{align*}
\end{theorem}

\medskip
The proof of the previous theorem is similar to that of the regret lower bound in \cite{lai1985}, and is omitted here. In view of this result, if we do not exploit structural properties of the problem, then the regret of any algorithm scales at least as $CK\log(T)$. Hence, when the number of channels and rates grow large, no algorithm is able to learn the best (channel, rate) pair rapidly and efficiently. 

\subsection{Structured MAB $(P_U)$}

To derive a regret lower bound for MAB problem $(P_U)$, we need to introduce additional notations. We define $M=N\cap \{k^\star-1, k^\star+1\}$. For any channel $c$, let $N_c=\{k: \mu_c^\star\le r_k\}$, and $k_{0c}$ such that $N_c=\{k_{0c},\ldots,K\}$, with the convention $k_{0c}=K+1$ if $N_c=\emptyset$. Observe that for any $c\neq c^\star$, $k_{0c}\le k_0$. Let $M_c=N_c\cap \{k_c^\star-1,k_c^\star+1\}$. The following theorem is proved in \cite{combesTR2014}. Due to space limitations, all the proofs are presented in \cite{combesTR2014}.

\begin{theorem}\label{th:lower_dep}
Let $\pi\in \Pi$ be a uniformly good (channel, rate) selection algorithm for MAB problem ($P_U$). We have: $\liminf_{T\to\infty} {R^\pi(T)\over \log(T)} \ge c_U({\theta})$,
where $c_U(\theta)$ is the optimal value of the following optimization problem:
\begin{align*}
&\inf_{\alpha_{ck}\ge 0,\forall k,c} \sum_{(c,k)\neq (c^\star,k^\star)} \alpha_{ck}( \mu^\star -\mu_{ck})\\
& \hbox{s.t. } \forall k\in M, \alpha_{c^\star k}I(\theta_{c^\star k},{\mu^\star \over r_k})\ge 1,\\
&\quad\forall c\neq c^\star: k_c^\star\ge k_0, \alpha_{ck_c^\star}I(\theta_{ck_c^\star},{  \mu^\star \over r_{k_c^\star}})\ge 1,\hbox{ and }\\
&\quad \forall k\ge k_0, k\neq k_c^\star, \inf_{\lambda_c\in C_k}\sum_{l}\alpha_{cl}I(\theta_{cl},\lambda_{cl})\ge 1,
\end{align*}
where $C_k=\{ \lambda_c\in {\cal U}\cap{\cal T}: r_k\lambda_{ck}>\mu^\star\}$.
\end{theorem}

\medskip
The above theorem does not provide a fully explicit regret lower bound. In particular, it remains unclear how this lower bound scales with the numbers of rates and channels. In the following theorem, we further exploit the structural properties of the MAB problem $(P_U)$ to show that $c_U(\theta)$ scales at most linearly with the number of channels, and does not scale with the number of rates. 

\begin{theorem}\label{th:lower_dep_scaling}
We have $c_U(\theta) \leq c_U'(\theta)$ where
\begin{align}\label{eq:lower_dep_scaling}
c_U'(\theta) =&  \sum_{k \in M} \frac{  \mu^\star - \mu_{c^\star k} }{ I( \theta_{c^\star k}, {\mu^\star\over r_{k}})} \nonumber\\
+& \sum_{c \neq c^\star} \bigg[ \frac{ \mu^\star  - \mu_{c k^\star_c} }{ \min \{ I( \theta_{c k^\star_c},  {\mu^\star \over r_{ k^\star_c}} ) ,  I( \theta_{c k^\star_c} ,  \theta_{c k^\star_c} - {\delta_c\over r_{ k^\star_c}}) \}} \nonumber\\
&+ \sum_{ k \in M_c} \frac{ \mu^\star - \mu_{c k}}{ I( \theta_{c k},  \theta_{c k} + {\delta_c\over r_k} )} \bigg].
\end{align}
and
\eqs{
\delta_c  =  \min_{ k \in \{k_c^\star-1,k_c^\star+1\} } ( \mu_{c k_c^\star} - \mu_{c k} ) / 2. 
}
In particular, $c_U'(\theta)$ is proportional to the number of channels and independent of the number of rates. 
\end{theorem}

From the above analysis, we conclude that the minimum regret for the MAB problem $(P_U)$ scales at most as $3C\log(T)$. Hence we expect that exploiting the structure of the problem (the fact that $\theta_c\in {\cal T}\cap{\cal U}$ for any channel $c$) may significantly improve the system performance. Indeed we expect a regret that does not depend on the number of available rates. In the next section, we design an algorithm with such a regret.

\subsection{Structured MAB problem $(P_{GU})$}

Graphical unimodal bandit problems have been recently studied in \cite{yu2011, combes2014}. A regret lower bound is derived in \cite{combes2014}. The only difference between our graphically unimodal MAB problem and those considered in  \cite{combes2014} is that we consider directed graphs, but the analysis is similar. We use here the notation introduced in \textsection \ref{not1}, and recall that $N=\{k:  \mu^\star \le r_k\}$. For any $(c,k)$, we define ${\cal N}'(c,k)={\cal N}(c,k)\cap N$. ${\cal N}'(c,k)$ is the set of (channel, rate) pairs that are neighbors of vertex $(c,k)$, and that need to be explored if one wants to know whether they provide better throughput than $(c,k)$.

\begin{theorem}\label{th:combes}\cite{combes2014}
Let $\pi\in \Pi$ be a uniformly good (channel, rate) selection algorithm for MAB problem ($P_{GU}$). We have:
\eq{
\lim \inf_{T \to +\infty} \frac{R^{\pi}(T)}{ \log(T)}  \geq c_{GU}(\theta), 
}
where 
$$
c_{GU}(\theta)= \sum_{(c,k)\in {\cal N}'(c^\star,k^\star)} \frac{ \mu^{\star} - \mu_{ck}}{I(\theta_{ck}, {\mu^{\star}\over r_k})}.
$$
\end{theorem}

\medskip
In view of the above theorem, for the MAB problem $(P_{GU})$, the minimum regret scales as $\gamma\log(T)$, where $\gamma$ is $\gamma$ is the maximum node degree in the graph $G$. Note that for our graph $G$, $\gamma\le 2C$. Hence, by exploiting the graphical unimodal structure, we may expect to design algorithms whose regret does not depend on the number of available rates. In the next section, an algorithm whose regret matches the lower bound of Theorem \ref{th:combes} is proposed.  

In this section, we have shown that the regret lower bound can be significantly improved when structural assumptions are made, i.e., $c_{GU}(\theta)\le c_U(\theta)\le c_I(\theta)$. By exploiting the structure of the problem, we may actually design algorithms whose regrets does not depend on the number of available rates. Such algorithms do not exist when the structure is not exploited (see Theorem \ref{th:lower_indep}).

\section{Algorithms}\label{sec:algo}

In this section, we present algorithms for the three MAB problems $(P_I)$, $(P_U)$, and $(P_{GU})$, and analyze their regrets. For the two structured MAB problems, the proposed algorithms exhibit a regret that does not depend on the number of available rates.

\subsection{The KL-UCB algorithm for MAB problem $(P_I)$}

Classical unstructured bandit problems have been extensively studied in the past, and numerous efficient algorithms have been proposed. We build on this previous work, and present a simple extension of KL-UCB algorithm \cite{garivier2011} to the MAB problem $(P_I)$. This algorithm does not exploit any structural properties, and is asymptotically optimal: its regret matches the lower bound derived in Theorem \ref{th:lower_indep}.

Under the KL-UCB algorithm, each (channel, rate) pair $(c,k)$ is associated with an index $q_{ck}(n)$ for the $(n+1)$-th packet transmission:
\begin{align*}
q_{ck}(n) =& \max \{q\in [0,r_k]: \\
& t_{ck}(n)I({\hat{\mu}_{ck}(n)\over r_k},{q\over r_k})\le \log(n)+3\log\log(n)\},
\end{align*}
where $t_{ck}(n)$ denotes the number of times $(c,k)$ has been selected up to the $n$-th transmission, and 
\eqs{
\hat\mu_{ck}(n) = \frac{1}{t_{ck}(n)}\sum_{i=1}^{t_{ck}(n)}  r_kX_{ck}(i),
}
is the empirical throughput or reward of (channel, rate) pair $(c,k)$ up to the $n$-th transmission. The algorithm selects the (channel, rate) pair with highest index:
\begin{separation}
\vspace{-0.1cm}
    {\bf Algorithm 1} KL-UCB
\vspace{-0.5cm}\separator
\vspace{-0.2cm}
For $n = 0,\dots,CK-1$ (initialization): for the $(n+1)$-th transmission, select (channel, rate) pair $(c,k)(n+1)=(c'+1, k'+1)$ where $n=Kc'+k'$, $k'\in\{0,\ldots,K-1\}$.
\newline For $n \ge CK$, for the $(n+1)$-th transmission, select $(c,k)(n+1)$ where $ (c,k)(n+1) \in \arg\max_{(c,k)} q_{ck}(n)$.
\vspace{-0.1cm}
\end{separation}

The rationale behind the design of KL-UCB is the same as that of the classical UCB algorithm. We construct an index for each (channel, rate) pair, which in turn constitutes an upper confidence bound of the corresponding throughput. By selecting the pair with the highest upper confidence bound, we force the exploration of suboptimal pairs if the latter have not been explored enough (in such a case, the upper confidence bound of a suboptimal pair can be higher than that of the optimal pair). KL-UCB is designed so that the number of times a suboptimal pair is selected matches the optimal number of times it is explored in the regret lower bound. In fact, KL-UCB is known to be asymptotically optimal in classical bandit problems \cite{garivier2011}. It can be easily established that its extension is also optimal for the problem $(P_I)$:

\medskip
\begin{theorem}\label{th:kl}
For any $\theta\in [0,1]^{C\times K}$, the regret of the $\pi= $ KL-UCB algorithm satisfies:
$$
\lim\sup_{T\to\infty}{R^{\pi}(T)\over \log(T)}\le c_I(\theta),
$$
\end{theorem}

\medskip
In particular, the regret under KL-UCB scales linearly with the numbers of channels and rates. When the later become large, the performance of KL-UCB can be quite poor.

Note that one may actually derive finite-time upper bounds on the regret of KL-UCB, as done in \cite{garivier2011}. KL-UCB is asymptotically optimal, but also provides good performance over a finite time horizon.

Finally, regarding the computational complexity of implementing KL-UCB, note that we just require to maintain an index for each pair, which requires a number of operations that scales as $CK$ after each transmission. The comparison between the various indexes can be done with $CK\log(CK)$ operations.
 
\subsection{The \ouralgo algorithm for MAB problem $(P_U)$}

Next, we present CRS-T (Channel and Rate Sampling with Tests), an algorithm that exploits the structure of the MAB problem $(P_U)$, i.e., the fact that on each channel, the throughput is a unimodal function of the rate. To describe our algorithm, we introduce the following notations. After the $n$-th transmission, the rate with the highest average empirical throughput on channel $c$ is referred to as the {\it leader} on channel $c$, and is $l_c(n) = \arg\max_{k} \hat\mu_{ck}(n)$. The {\it global leader} $l(n)$ is the (channel, rate) pair with highest average empirical throughput: $l(n) =  \arg\max_{(c,k)} \hat\mu_{ck}(n)$.

Similar to the KL-UCB index $q_{ck}(n)$, we define the lower confidence bound $\underline{q}_{ck}(n)$ as:
\begin{align*}
\underline{q}_{ck}(n) =& \min \{q\in [0,r_k]: \\
& t_{ck}(n)I({\hat{\mu}_{ck}(n) \over r_k},{q\over r_k}) \le \log(n)+3\log\log(n)\},
\end{align*}

We introduce a statistical test, which will be used to assess whether the leader on channel $c$, $l_c(n)$, provides a larger reward than its neighbors $l_c(n)-1$, $l_c(n)+1$ on the same channel. Define the test for channel $c$ at time $n$ through $U_{c}(n):$
\als{
	U_{c}(n) = \indic \{  \underline{q}_{c l_c(n)}(n) \geq \sup_{k:|k - l_c(n)| = 1}  q_{ck}(n) \}
}

The test can be interpreted as follows. $U_{c}(n) = 1$ means that $l_c(n)$ is better than its neighbors with high probability, and $U_{c}(n) = 0$ means that we do not have enough samples to determine whether $l_c(n)$ is better than its neighbors. After the $n$-th packet transmission, we define ${\cal U}(n) = \{ c:  U_{c}(n) = 0 \}$ the set of channels for which we cannot determine whether the leader $l_c(n)$ corresponds to the best rate $k^\star_c$ on this channel. 

The sequential decisions under the CRS-T algorithm are based on the indexes of the various (channel, rate) pairs, and can be easily implemented. The index $b_k(n)$ of decision $(c,k)$ for the $(n+1)$-th packet transmission is:
\eqs{
b_{ck}(n) = q_{ck}(n)  \indic\{ k = l_c(n) \},
}
where $q_{ck}(n)$ is the index used in the KL-UCB algorithm. Note that the index of decision $(c,k)$ is equal to $0$ if $k$ is not the leader on channel $c$. The pseudo-code for CRS-T is given below (each time the decision is ambiguous, ties are broken arbitrarily). 

\begin{separation}
\vspace{-0.1cm}
    {\bf Algorithm 2} CRS-T
\vspace{-0.6cm}\separator
\vspace{-0.3cm}
For $n = 0,\dots,CK-1$ (initialization): for the $(n+1)$-th transmission, select (channel, rate) pair $(c,k)(n+1)=(c'+1, k'+1)$ where $n=Kc'+k'$, $k'\in\{0,\ldots,K-1\}$.
\newline For $n \ge CK$: for the $(n+1)$-th transmission, select $(c,k)(n+1)$ where
\begin{itemize}
\item if ${\cal U}(n) \neq \emptyset$, then $c(n+1) \in {\cal U}(n)$ and\\ $k(n+1) \in \arg \min_{k':|k' - l_{c(n+1)}(n)| \leq 1} t_{c(n+1)k'}(n)$;
\item else $(c,k)(n+1) \in \arg \max_{c',k'} b_{c'k'}(n)$. 
\end{itemize}
\vspace{-0.2cm}
\end{separation}

The design of the CRS-T algorithm is motivated by the following objectives:  (1) For all channels, we need to play the leader and all its neighbours until we can determine with high probability that the leader $l_c(n)$ is the best rate $k^\star_c$. (2) Once we have determined that $l_c(n)$ is $k^\star_c$ for all channels, then we play the leader of the channel with the largest index, i.e we apply KL-UCB restricted to the set of leaders.

Define $\Delta = \min_{(c,k)} \min_{k':|k'-k|=1 } | \mu_{ck} - \mu_{ck'}  |$ the minimal separation between two neighboring rates on any channel and $\tilde{\mu}_c = (\mu_{c k_c^\star} +  \max_{|k-k_c^\star|=1} \mu_{ck})/2$. The next theorem, proved in \cite{combesTR2014}, provides a finite time upper bound on the regret under CRS-T. The asymptotic regret $\log(T) c^{\text{CRS-T}}(\theta)$ scales linearly with the number of channels, but is independent of the number of available rates. In particular, CRS-T exploits the structure of the MAB problem $(P_U)$.

\medskip
\begin{theorem}\label{th:crs}
For any $\theta$ such that for all $c$, $\theta_c\in {\cal T}\cap{\cal U}$, and for all $\epsilon > 0$, the regret of the CRS-T algorithm satisfies:
\begin{align*}
  R^{\text{CRS-T}}(T) \leq (1+\epsilon)& c^{\text{CRS-T}}(\theta)\log(T) \\
  &+ \Gamma_{\epsilon} ( \Delta^{-2}+ \log(\log(T)) ),
\end{align*}
where $\Gamma_{\epsilon} > 0$ depends on $\epsilon$, $C$, $K$ and ${\cal R}$ but not on $\theta$, and
\als{
	& c^{\text{CRS-T}}(\theta) = \sum_{c=1}^{C} \tau_{c}^{-1} \sum_{k:|k - k_c^\star| \leq 1} (\mu^\star - \mu_{ck}),
}
with
\eqs{
\tau_{c} =   \min \left(  \min_{k:|k - k_c^\star| \leq 1} I(\theta_{ck}, \tilde{\mu}_c/r_k) , I(\theta_{ck_c^\star}, \mu^\star/r_{k_c^\star}) \right)
}
\end{theorem}

\medskip
The regret under CRS-T does not depend on the number of available rates, and hence exploits (at least asymptotically) the unimodal structure of $(P_U)$. The computational complexity of CRS-T is similar to that of KL-UCB because it essentially requires to maintain the indexes of the various channel and rate pairs: it scales linearly with $CK$ (up to a logarithmic factor).

\subsection{The KL-UCB-U algorithm for MAB problem $(P_{GU})$}

Finally, we present KL-UCB-U, an algorithm for MAB problem $(P_{GU})$. KL-UCB-U is a natural extension of an algorithm proposed in \cite{combes2014} for graphically unimodal bandits with undirected graphs. This algorithm is asymptotically optimal (its regret matches the lower bound derived in Theorem \ref{th:combes}).

Recall that the global leader is denoted by $l(n)$ before the $(n+1)$-th transmission. We introduce $v_{(c,k)}(n)$ the number of times that (channel, rate) pair $(c,k)$ has been the global leader up to the $n$-th transmission: $v_{(c,k)}(n)=\sum_{n'=1}^n \indic{ \{ l(n')=(c,k)}\}$. The index associated with decision $(c,k)$ before the $(n+1)$-th transmission is:
\begin{align*}
b_{ck}(n)= & \max \Big\{ q \in [0,r_k] : t_{ck}(n)I\big(\frac{\hat{\mu}_{ck}(n)}{r_k}, \frac{q}{r_k} \big) \\
&\leq \log (v_{l(n)}(n))+3\log(\log (v_{l(n)}(n))) \Big\},
\end{align*}
For the $(n+1)$-th transmission, KL-UCB-U selects the (channel, rate) pair in the neighborhood of the leader with maximum index. Ties are broken arbitrarily. 

\begin{separation}
\vspace{-0.1cm}
    {\bf Algorithm 3} KL-UCB-U
\vspace{-0.5cm}\separator
\vspace{-0.2cm}
For $n = 0,\dots,CK-1$ (initialization): for the $(n+1)$-th transmission, select (channel, rate) pair $(c,k)(n+1)=(c'+1, k'+1)$ where $n=Kc'+k'$, $k'\in\{0,\ldots,K-1\}$.
\newline For $n \ge CK$: for the $(n+1)$-th transmission, select $(c,k)(n+1)$ where:
\vspace{0.2cm}
 $$ (c,k)(n+1) = \begin{cases}  l(n)  \quad \text{ if } (v_{l(n)}(n) - 1)/\gamma \in \mathbb{N}, \\
	  \displaystyle \arg\max_{(c,k) \in {\cal N}(l(n))} b_{ck}(n) \quad \text{ otherwise.}
	  \end{cases} $$
\vspace{-0.1cm}
\end{separation}

Remember that $\gamma$ is the maximum number neighbors in $G$ of a given  (channel, rate) pair. The KL-UCB-U algorithm periodically selects the leader to make sure that the latter is often selected. The design of KL-UCB-U is based on the lower regret bound derived for the MAB problem $(P_{GU})$. This lower bound implies that an optimal algorithm explores suboptimal (channel, rate) pairs a number of times that scales with $\log(T)$ only for pairs that are neighbours of the the optimal pair in the graph $G$. Hence in KL-UCB-U, the exploration is restricted to the neighbours of the current leader in $G$. As in \cite{combes2014}, we can establish that KL-UCB-U is asymptotically optimal:

\medskip
\begin{theorem}
For any $\theta\in {\cal T}^C\cap{\cal U}_G$, the regret of $\pi= $KL-UCB-U satisfies:
$$
\lim\sup_{T\to\infty}{R^\pi(T)\over \log(T)}\le c_{GU}(\theta),
$$
\end{theorem}

\medskip
In particular, KL-UCB-U optimally exploits the structure of MAB problem $(P_{GU})$. In turn, if the throughput is a graphically unimodal function of the (channel, rate) pair, then KL-UCB-U asymptotically outperforms any other algorithm, and in particular \ouralg, an algorithm designed to exploit the unimodal structure per channel only.

Note that a finite-time regret analysis for KL-UCB-U is possible as shown in \cite{combes2014}. KL-UCB-U is asymptotically optimal, but also provides good performance over a finite time horizon. Finally, again, the computational complexity of KL-UCB-U is similar to that of KL-UCB.

\section{Non-stationary Radio Environments}\label{sec:nonstat}

In practice, channel conditions may be non-stationary, i.e., the success probabilities at various (channel, rate) pair could evolve over time. In many situations, the evolution over time is rather slow -- refer to \cite{radunovic2011} and to Section V for test-bed measurements. These slow variations allow us to devise (channel,rate) adaptation schemes that efficiently track the best (channel,rate) pair for transmission. 

We assume that for all $(c,k)$ pairs, the transmissions outcomes $X_{ck}(n)$ , $n=1,2,\dots$ are independent, with expectation $\theta_{ck}(n)= \EE[X_{ck}(n)]$. At time $n$ we define the throughput of $(c,k)$ $\mu_{ck}(n) = r_k \theta_{ck}(n)$, the best throughput $\mu^\star(n) = \max_{c,k} \mu_{ck}(n)$ and the optimal decision $(c^\star,r^\star)(n) = \arg \max_{c,k} \mu_{ck}(n)$.
	
Any algorithm designed for stationary radio environments can readily be extended to non-stationary environments. These extensions are obtained by replacing empirical averages by averages over a sliding time window. Let $\tau \geq 1$ denote the sliding window size, and define the empirical reward $\hat\mu_{ck}(n)$ as:
$$
\hat\mu^\tau_{ck}(n) = { r_k \over t^\tau_{ck}(n) } \sum_{n'=n - \tau + 1}^n   X_{ck}(n') \indic\{(c,k)(n') = (c,k)\},
$$
where
$$
t^\tau_{ck}(n) = \sum_{n'=n - \tau + 1 }^n \indic\{ (c,k)(n') = (c,k)\}, \sk
$$	
with the convention $\hat\mu^\tau_{ck}(n) = 0$ if $t^\tau_{ck}(n) = 0$. We also define the upper confidence index of (channel, rate) pair $(c,k)$ as:
\als{
q^\tau_{ck}&(n) = \max \{q \in [0,r_k] :\sk 
&  I({\hat\mu^\tau_{ck}(n)\over r_k} , {q\over r_k} ) \leq \log(\tau) + 3 \log(\log(\tau))  \}.
}

We define sliding window variants of the algorithms presented in Section~\ref{sec:algo} by replacing $t_{ck}(n)$ by $t^\tau_{ck}(n)$, $\hat\mu_{ck}(n)$ by $\hat\mu^\tau_{ck}(n)$ and $q_{ck}(n)$ by $q^\tau_{ck}(n)$. For instance, KL-UCB with sliding window is the algorithm which selects $(c,k)(n) \in \arg \max_{ck} q^\tau_{ck}(n)$.

In \cite{garivier2008}, the authors show that algorithms with sliding windows efficiently track the best decision over time provided that the environment evolves relatively slowly. This is confirmed in \cite{combes2014}, where the performance of algorithms similar to KL-UCB and KL-UCB-U with sliding window is analyzed. Due to space limitation, we skip this analysis; refer to \cite{combes2014} for more details.

The way the window size $\tau$ should be chosen in practice is dictated by the following remarks. First, $\tau$ should be relatively small compared to the time it takes for the packet successful transmission probabilities to evolve. This ensures that the algorithms with sliding window track the best channel and rate pair. Then, $\tau$ should be sufficiently large so that the throughput of the various channel and rate pairs could be estimated with precision using samples collected in an interval of time of duration $\tau$. Typically 10 or 20 packets sent using the same channel and rate pair is enough. These two requirements for $\tau$ go in opposite directions, and clearly when the packet successful transmission probabilities evolve rapidly, an appropriate design of $\tau$ is not possible. In practice, $\tau$ may be tuned by just observing the way these probabilities evolve (via off-line preliminary experiments). It is not necessary to adapt $\tau$ over time if the radio environment does not change (say for example, users stay inside a building). However, in other scenarios, we may need to adapt $\tau$.

\section{Numerical Experiments}\label{sec:num}
In this section we numerically illustrate the performance of the proposed (channel, rate) selection algorithms. To this aim, we conduct simulation experiments where our algorithms are tested against channel quality traces that are either extracted from a test-bed \cite{radunovic2011}, or artificially generated. In the latter case, we generate traces based on a widely used statistical model for radio propagation and on a mapping between channel quality and probability of packet successful transmission on a given (channel,rate) pair \cite{halperin2010}. 

\subsection{Traces extracted form a test-bed}\label{ssec:testbed}

In this subsection we present trace-driven experiments using the test-bed described in~\cite{radunovic2011}. The test-bed is based on a SDR platform (Lyrtech SFF-SDR), and is located in an indoor office. The PHY layer is OFDM, as in 802.11a/g/n. There are $3$ available rates $\{4.5,6,6.75\}$ Mbps corresponding to QPSK modulation with respective coding rates $\{1/2,2/3,3/4 \}$. We consider $5$ channels in the UHF band centred at $\{510,530,550,580,600\}$ Mhz . The bandwidth of each channel is $10$ Mhz, and the packet size is $1500$ bytes. The trace duration is $600$s. 

The traces are collected as follows. The transmitter transmits 10 packets at each rate on a given channel before moving to the next channel. Each measurement round (where each (channel, rate) pair is probed) hence consists of the transmission of $10\times 5\times 3$ packets, and lasts less than 1s. This means we sample
a given channel at a given rate once every second. In each round and for each channel, we calculate, by averaging over 10 packets, the RSSI, the successful packet transmission probabilities and the goodputs at the 3 different rates.

\begin{figure}
	\includegraphics[width=1\columnwidth]{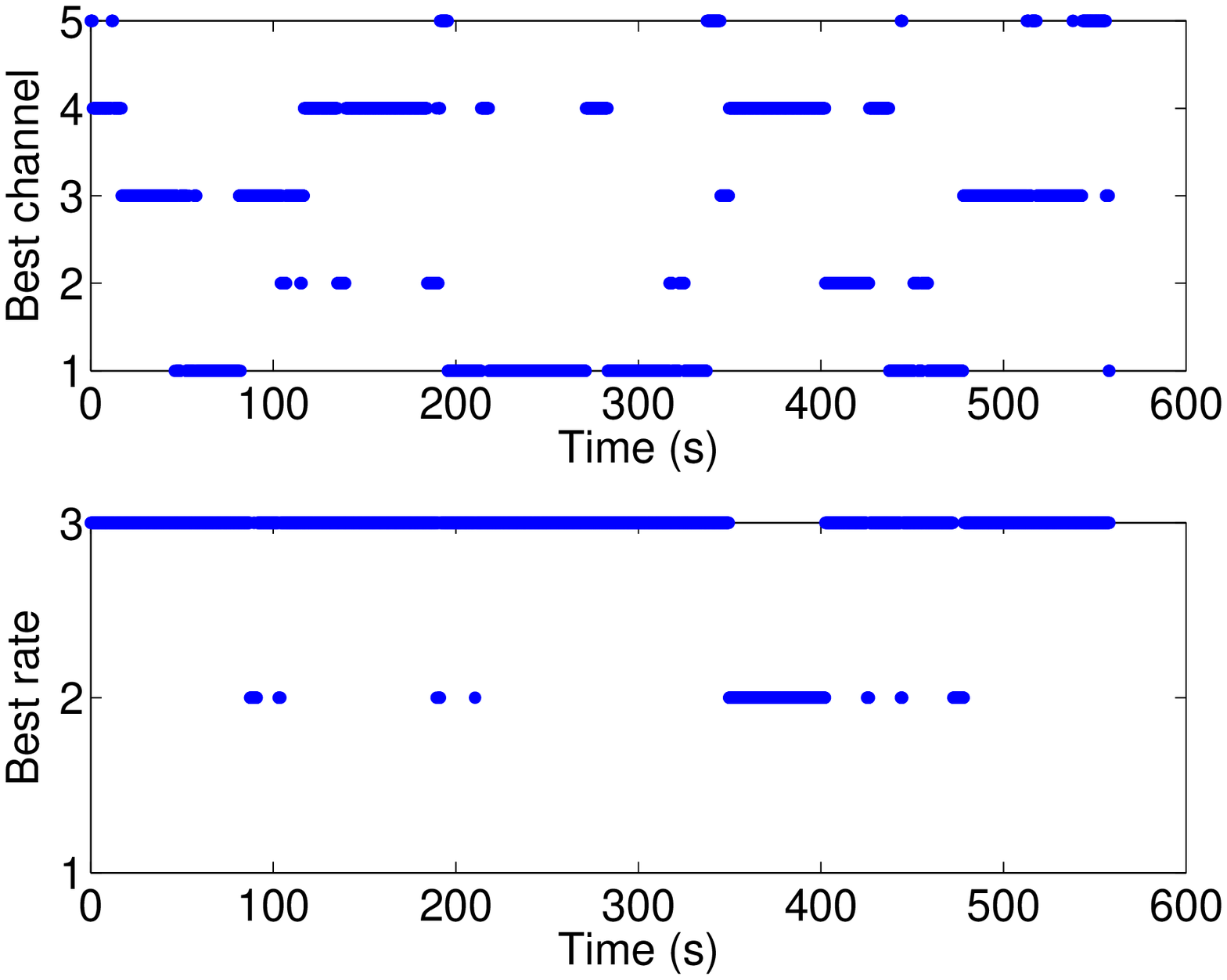}
	\vspace{-10mm}
	\caption{Test-bed: best (channel, rate) pair $(c^\star,k^\star)(t)$ as a function of time.}
	\label{fig:real_trace_best_decision}
\end{figure}

In Fig.\ref{fig:real_trace_best_decision}, we plot the best decision $(c^\star,k^\star)$ as a function of time. the radio environment is non-stationary, and the optimal decision remains constant for several seconds. Since a packet transmission lasts about $1$ms, the packet successful transmission probabilities for various decisions stay constant for thousands of packet transmissions. Therefore we have quite a lot of statistical information to find the best decision. Furthermore the window size used in the tested algorithms should be of the order of a few seconds -- we fix it to 2s.

\begin{figure}
	\includegraphics[width=1\columnwidth]{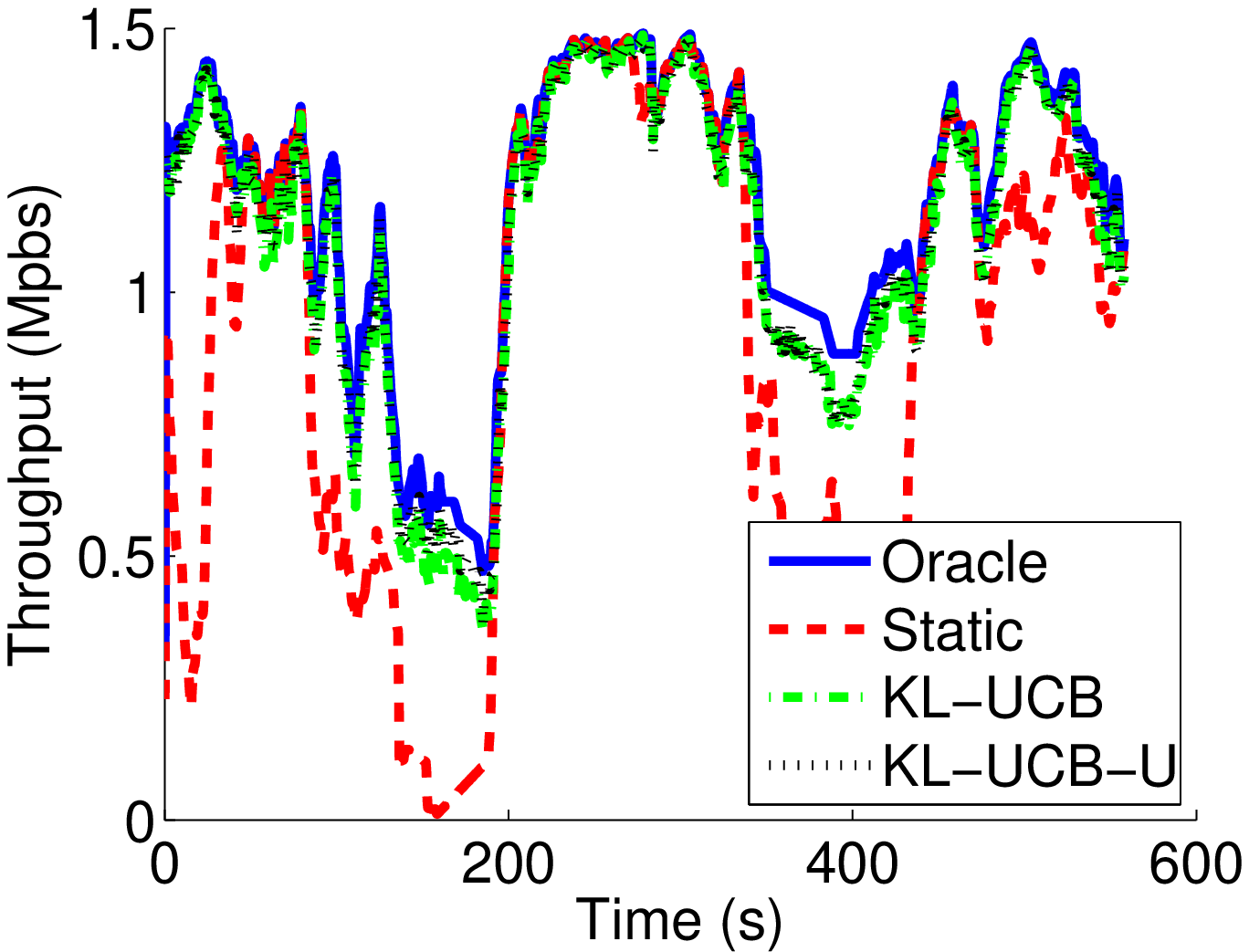}
	\vspace{-10mm}
	\caption{Test-bed: throughputs of the various algorithms as a function of time.}
	\label{fig:real_trace}
\end{figure}

In Fig.\ref{fig:real_trace}, we plot the throughput under KL-UCB and KL-UCB-U algorithms. For the sake of comparison, we also plot $\mu^\star(t)$ the throughput of an Oracle algorithm that always selects the optimal decision. We also plot the throughput obtained by choosing the best static (channel, rate) pair, computed offline. We observe that selecting the best static pair is clearly sub-optimal, so that adaptive algorithms can lead to a large gain in throughput. Both decision algorithms, KL-UCB and KL-UCB-U, manage to closely follow the best (channel, rate) pair. KL-UCB-U provides a throughput equal to $95\%$ of that obtained under the Oracle algorithm, whereas the throughput under KL-UCB is equal to $90\%$ of that of the Oracle algorithm. There is not a huge performance gap between KL-UCB and KL-UCB-U because there are few available rates,  $K=3$. Hence KL-UCB explores $C \times K = 15$ (channel,rate) pairs, while KL-UCB-U explores (in the worse case) $2 C + 1 = 11$ pairs. We will show that increasing the number of available rates $K$ makes this difference significantly larger. 
 
\subsection{Artificial traces}\label{ssec:simulation}

We also present numerical results based on a widely used statistical model for radio propagation. Namely, we assume that the channel is a multi-path Rayleigh fading channel. When a signal is transmitted, several delayed copies of this signal are received and the amplitude and phase of each delayed copy is an independent Rayleigh fading process. We use Jakes' model to simulate Rayleigh fading with user speed set to match the time variability of the test-bed trace presented in~\ref{ssec:testbed}. This corresponds to static users such as laptops in an office environment. The expected power of each delayed path is chosen according to the field measurements presented in ~\cite{radunovic2013}. 

We assume that OFDM is used, and the mapping between the strength of received signal on each sub-carrier and the probability of successful transmission is calculated by the method presented in~\cite{halperin2010}. We consider $5$ channels with bandwidth $20$ Mhz in the $2.4$ GHz band centred at $\{2.4,2.41,2.42,2.43,2.44\}$GHz, respectively. Each channel has $52$ sub-carriers and the packet size is $1500$ bytes. We consider $8$ available rates: $\{6,13,19.5,26,39,52,58.5,65\}$ Mbps, and a transmitter-receiver pair with an average SNR of $20$ dB. The trace length is $600$ seconds.

We first consider stationary environments, so that a snapshot of the success probabilities for all (channel,rate) pairs is drawn and kept constant throughout the simulation. Fig.\ref{table:stationary_trace_mu} shows the packet successful transmission probabilities and throughputs of different (channel,rate) pairs. As announced, graphical unimodality holds: the throughput on each channel is a unimodal function of the rate, and given the optimal rate $k_c^\star$ on sub-optimal channel $c \neq c^\star$, there exists another channel $c' \neq c$ such that either $\mu_{c',k_c^\star} > \mu_{c,k_c^\star}$ or $\mu_{c',k_c^\star+1} > \mu_{c,k_c^\star}$. Graphical unimodality results from the fact that we are in a steep environment as defined in \cite{bicket2005}. Fig.\ref{fig:stationary_trace} presents the regret of KL-UCB and KL-UCB-U as a function of time. KL-UCB-U beats KL-UCB and for large time horizons the regret under KL-UCB-U is roughly half of that under KL-UCB. Hence exploiting the graphical unimodal structure significantly helps.
\begin{figure}
\begin{center}
\begin{tabular}{|c|c|c|c|c|c|c|c|c|}
\hline 
$r_k$ & 6 & 13 & 19.5 & 26 & 39 & 52 & 58.5 & 65 \\ 
\hline 
$\theta_{1,k}$  & 1 & 1 & 1 & 1 & 1 & 0.2 & 0 & 0 \\ 
$\theta_{2,k}$  & 1 & 1 & 1 & 1 & 1 & 1 & 0.7 & 0.1 \\ 
$\theta_{3,k}$  & 1 & 1 & 1 & 1 & 1 & 0.6 & 0 & 0 \\ 
$\theta_{4,k}$  & 0 & 0 & 0 & 0 & 0 & 0 & 0 & 0 \\ 
$\theta_{5,k}$  & 1 & 1 & 0.8 & 0.2 & 0 & 0 & 0 & 0 \\ 
\hline 
$\mu_{1,k}$ & 6 & 13 & 19.5 & 26 & 39 & 13 & 0 & 0 \\ 
$\mu_{2,k}$ & 6 & 13 & 19.5 & 26 & 39 & 52 & 41 & 8 \\ 
$\mu_{3,k}$ & 6 & 13 & 19.5 & 26 & 39 & 29 & 0 & 0 \\ 
$\mu_{4,k}$ & 0 & 0 & 0 & 0 & 0 & 0 & 0 & 0 \\ 
$\mu_{5,k}$ & 6 & 13 & 16 & 6 & 0 & 0 & 0 & 0 \\ 
\hline 
\end{tabular}
\end{center}
\caption{Simulation: Packet successful transmission probabilities and throughputs at different (channel,rate) pairs in a stationary environment.}
\label{table:stationary_trace_mu}
\end{figure}
\begin{figure}
	\includegraphics[width=1\columnwidth]{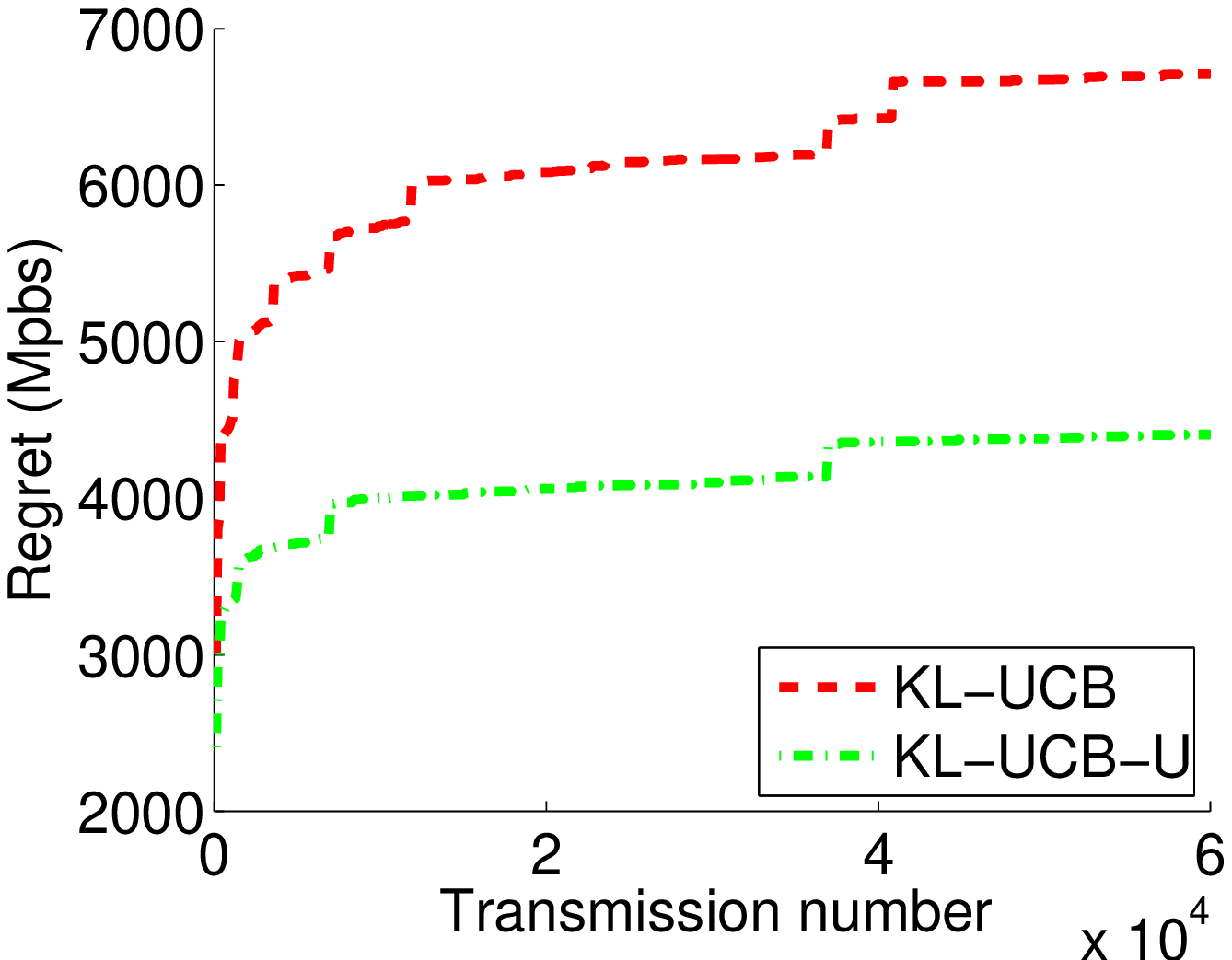}
		\vspace{-10mm}
	\caption{Simulation: Regret of different decision rules as a function of time in a stationary environment.}
	\label{fig:stationary_trace}
\end{figure}
	
\begin{figure}
	\includegraphics[width=1\columnwidth]{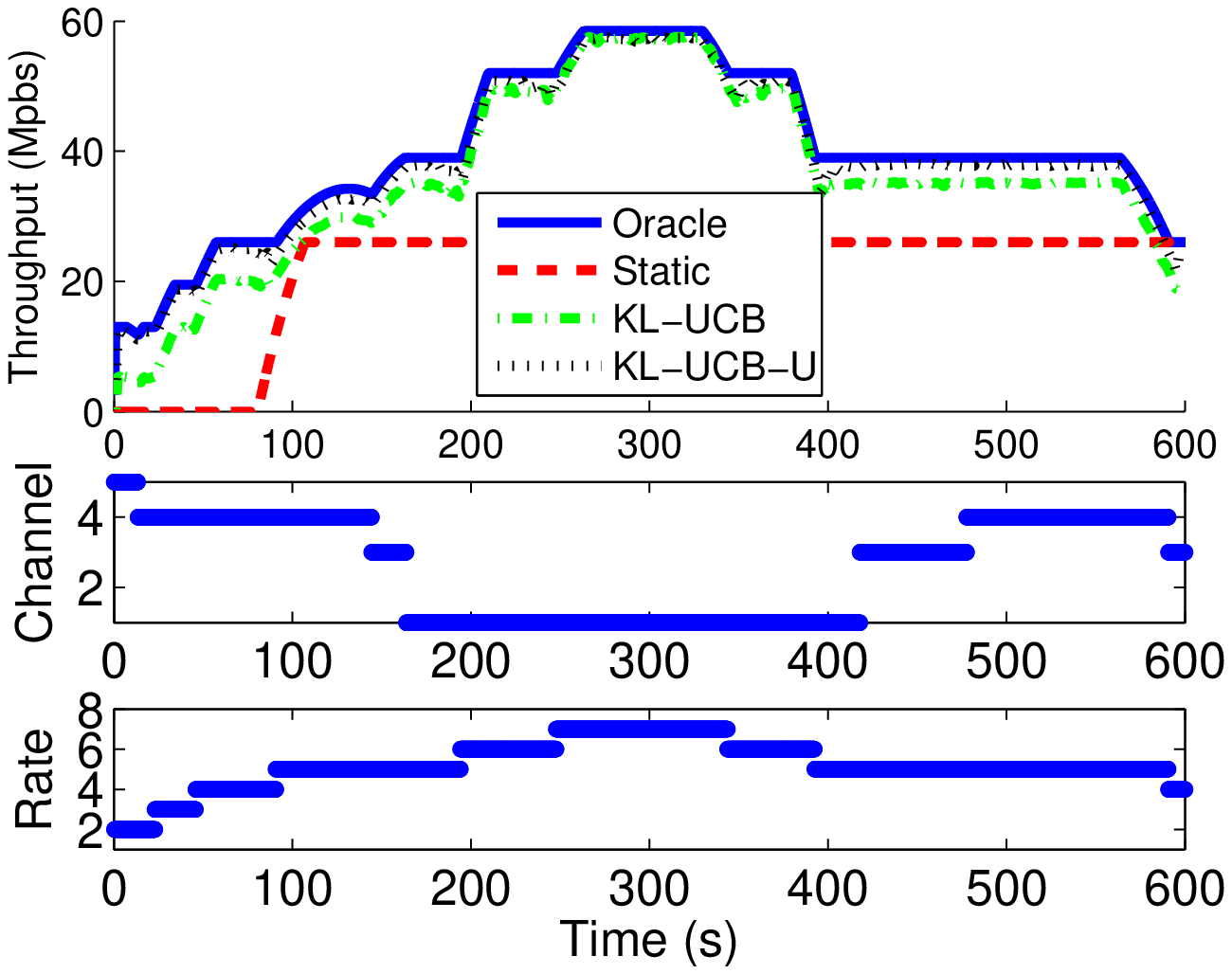}
		\vspace{-10mm}
	\caption{Simulation: throughputs of the various algorithms (above) and the best pair $(c^\star,k^\star)$ (below) as a function of time in a non-stationary environment -- low variation speed.}
	\label{fig:arti_trace_all}
\end{figure}

We now turn to non-stationary environments. As in~\ref{ssec:testbed}, we present the best pair as a function of time and the throughputs of different algorithms in Fig.\ref{fig:arti_trace_all}. Again KL-UCB-U beats KL-UCB.

For both the test-bed and simulation, the performance of KL-UCB-U is rather impressive: its throughput is at least $95\%$ of that of the Oracle, without knowing the throughputs of the various (channel,rate) pairs beforehand. This shows that given a good decision rule, the selection of channel and rate can be done solely based on ACK/NACK feedback with excellent performance. This is critical for real-world systems because feedback of channel measurements is problematic in practice both in terms of delay and overhead. 

So far, in non-stationary environments, the packet successful transmission probabilities were evolving slowly. Next we vary the speed at which they evolve, by artificially accelerating our traces by a factor 20 and 100. Results are presented in Fig. \ref{table:efficiency_oracle}. At all speeds, KL-UCB-U beats KL-UCB, and the performance gap between the two algorithms increases with the speed. When the environment changes faster, the performance of KL-UCB becomes poor, as the algorithm needs to explore all (channel, rate) pairs, and cannot track the best pair. On the contrary, KL-UCB-U exploits the structure and explores less, which makes its performance more robust. 

\begin{figure}
\begin{center}
\begin{tabular}{cccc}
\hline
Speed & $\times 1$ & $\times 20$ &  $\times 100$ \\
\hline
Static 		&  52 \%   & 45 \%   &  43 \% \\
KL-UCB 		&  90 \%   & 83 \%   &  57 \% \\
KL-UCB-U 	&  96 \%   & 91 \%   &  79 \% \\
Oracle 		&  100 \% & 100 \%  &  100 \% \\
\hline
\end{tabular}
\end{center}
\caption{Impact of the speed of variation of the successful transmission probabilities on performance in a non-stationary environment.}
\label{table:efficiency_oracle}
\end{figure}
\section{Conclusion}\label{sec:concl}

In this paper, we have addressed the problem of joint channel and rate adaptation in cognitive radio systems. We have shown that the problem is equivalent to a structured MAB problem, where the structure stems from inherent properties of the throughput as a function of the selected channel and rate. For several assumptions on this structure, we have derived fundamental performance limits satisfied by any sequential (channel, rate) selection algorithm. For each structure type, we have also proposed algorithms which are either close or achieve these limits. Finally we have assessed the efficiency of the proposed algorithms through trace-driven experiments and simulations. The two key insights from our results are: (a) The channel and rate adaptation problem has a strong structure. This structure can be exploited to devise algorithms whose performance does not depend on the number of available rates, and is close to that of an Oracle algorithm that perfectly knows the packet successful transmission probabilities at any available (channel, rate) pair. (b) There exist readily implementable algorithms which allow almost perfect channel and rate selection without the need of any measurement and explicit feedback of the quality of the various channels.

\bibliographystyle{IEEEtran}
\bibliography{RA}

\appendix

\section*{Proof of Theorem \ref{th:lower_dep} }

We derive here the regret lower bounds for the MAB problem $(P_U)$. To this aim, we apply the techniques used by Graves and Lai \cite{graves1997} to investigate efficient adaptive decision rules in controlled Markov chains. We recall here their general framework. Consider a controlled Markov chain $(X_t)_{t\ge 0}$ on a finite state space ${\cal S}$ with a control set $U$. The transition probabilities given control $u\in U$ are parametrized by $\theta$ taking values in a compact metric space $\Theta$: the probability to move from state $x$ to state $y$ given the control $u$ and the parameter $\theta$ is $p(x,y;u,\theta)$. The parameter $\theta$ is not known. The decision maker is provided with a finite set of stationary control laws $G=\{g_1,\ldots,g_K\}$ where each control law $g_j$ is a mapping from ${\cal S}$ to $U$: when control law $g_j$ is applied in state $x$, the applied control is $u=g_j(x)$. It is assumed that if the decision maker always selects the same control law $g$ the Markov chain is then irreducible with stationary distribution $\pi_\theta^g$. Now the reward obtained when applying control $u$ in state $x$ is denoted by $r(x,u)$, so that the expected reward achieved under control law $g$ is: $\mu_\theta(g)=\sum_xr(x,g(x))\pi_\theta^g(x)$. There is an optimal control law given $\theta$ whose expected reward is denoted $\mu_\theta^{\star}\in \arg\max_{g\in G} \mu_\theta(g)$. Now the objective of the decision maker is to sequentially control laws so as to maximize the expected reward up to a given time horizon $T$. As for MAB problems, the performance of a decision scheme can be quantified through the notion of regret which compares the expected reward to that obtained by always applying the optimal control law.

We now apply the above framework to our MAB problem. For $(P_U)$, for all $c$, the parameter $\theta_c$ takes values in ${\cal T}\cap {\cal U}$. The Markov chain has values in ${\cal S}=\{0,r_1,\ldots,r_K\}$. The set of control laws is $G=\{1,\ldots,C\}\times \{1,\ldots,K\}$. These laws are constant, in the sense that the control applied by control law $(c,k)$ does not depend on the state of the Markov chain, and corresponds to selecting (channel, rate) pair $(c,k)$. The transition probabilities are given as follows: for all $x,y\in {\cal S}$, 
$$
p(x,y;(c,k),\theta)=p(y;(c,k),\theta)=\left\{
\begin{array}{ll}
\theta_{ck}, & \hbox{if }y=r_k,\\
1-\theta_{ck}, & \hbox{if }y=0.
\end{array}\right.
$$
Finally, the reward $r(x,(c,k))$ does not depend on the state and is equal to $r_k\theta_{ck}$, which is also the expected reward obtained by always using control law $(c,k)$. 

We now fix $\theta$: $\forall c$, $\theta_c\in {\cal T}\cap {\cal U}$. Define $I^{(c,k)}(\theta,\lambda)=I(\theta_{ck},\lambda_{ck})$ for any $(c,k)$. Further define the set $B(\theta)$ consisting of all {\it bad} parameters $\lambda$: $\forall c$, $\lambda_c\in {\cal T}\cap{\cal U}$ such that $(c^\star,k^{\star})$ is not optimal under parameter $\lambda$, but which are statistically {\it indistinguishable} from $\theta$:
\begin{align*}
B(\theta)=\{ \lambda:\forall c, & \lambda_c \in {\cal T}\cap {\cal U},\\
& \lambda_{c^\star k^{\star}} = \theta_{c^\star k^{\star}}, \max_{(c,k)}  r_k\lambda_{ck} > \mu^\star \},
\end{align*}
$B(\theta)$ can be written as the union of sets $B_{ck}(\theta)$ defined as:
$$
B_{ck}(\theta)=\{ \lambda \in B(\theta) :  r_k\lambda_{ck} > r_{k^{\star}}\lambda_{c^\star k^{\star}} \}.
$$
Note that $B_{ck}(\theta)=\emptyset$ if $r_k<r_{k^\star}\theta_{c^\star k^\star}$, hence if $k\notin N$. 

By applying Theorem 1 in \cite{graves1997}, we know that $c_U(\theta)$ is the minimal value of the following LP:
\begin {eqnarray}
\textrm{min }  & \sum_{c,k} \alpha_{ck}(\mu^\star - r_k\theta_{ck}) \\
\textrm{s.t. } & \inf_{\lambda \in B(\theta)} \sum_{(c,k) \neq (c^\star,k^{\star})} \alpha_{ck}I^{(c,k)}(\theta,\lambda) \geq 1, \label{eq:con1}\\
& \alpha_{ck} \geq 0, \quad \forall (c,k).
\end {eqnarray}

Next we detail the constraints (\ref{eq:con1}). These constraints are equivalent to:
\begin{align}
&\inf_{\lambda \in B_{c^\star k}(\theta)} \sum_{(c,l) \neq (c^\star,k^{\star})} \alpha_{cl}I^{(c,l)}(\theta,\lambda) \geq 1, \forall k\neq k^\star\label{eq:c1}\\
&\inf_{\lambda \in B_{c k_c^\star}(\theta)} \sum_{(c',l) \neq (c^\star,k^{\star})} \alpha_{c'l}I^{(c',l)}(\theta,\lambda) \geq 1, \forall c\neq c^\star\label{eq:c2}\\
&\inf_{\lambda \in B_{ck}(\theta)} \sum_{(c',l) \neq (c^\star,k^{\star})} \alpha_{c'l}I^{(c',l)}(\theta,\lambda) \geq 1, \forall c\neq c^\star,\forall k\neq k_c^\star\label{eq:c3}.
\end{align}

\noindent
\underline{Constraint (\ref{eq:c1}).} We prove that (\ref{eq:c1}) is equivalent to:
\begin{equation}\label{eq:con2}
\min_{k\in M} \alpha_{c^\star k}I(\theta_{c^\star k},{\mu^\star\over r_k}) \ge 1.
\end{equation}
Observe that if $k<k_0$ (i.e., if $k\notin N$), then $B_{c^\star k}(\theta)=\emptyset$. Let $k\in N$ with $k\neq k^\star$. Without loss of generality assume that $k>k^\star$. We prove that:
\begin{equation}\label{eq:inf}
\inf_{\lambda \in B_{c^\star k}(\theta)} \sum_{(c,l) \neq (c^\star,k^{\star})} \alpha_{c l}I^{(c l)}(\theta,\lambda) = \sum_{l=k^\star+1}^k \alpha_{c^\star l}I(\theta_{c^\star l},{\mu^\star\over r_l}).
\end{equation}
This is due to the fact that we can always choose $\lambda_{cl}=\theta_{cl}$ for all $c\neq c^\star$, and to the following two observations:
\begin{itemize}
\item for all $\lambda\in B_{c^\star k}(\theta)$, we have $\lambda_{c^\star k^\star}r_{k^\star}= \theta_{c^\star k^\star}r_{k^\star}$ and $\lambda_{c^\star k}r_{k}> \lambda_{c^\star k^\star}r_{k^\star}$, which using the unimodality of $\lambda$, implies that for any $l\in \{k^\star,\ldots,k\}$, $\lambda_{c^\star l}r_l\ge \theta_{c^\star k^\star}r_{k^\star}$. Hence:
$$
\sum_{l \neq k^{\star}} \alpha_{c^\star l}I^{(c^\star,l)}(\theta,\lambda) \ge \sum_{l=k^\star+1}^k \alpha_{c^\star l}I(\theta_l,{\mu^\star\over r_l}).
$$
\item For $\epsilon >0$, define $\lambda_\epsilon$ as follows: for all $l\in \{k^\star,\ldots,k\}$, $\lambda_{c^\star l} = (1+(l-k^\star)\epsilon) {\mu^\star\over r_l}$, and for all $l\notin \{k^\star,\ldots,k\}$, $\lambda_{c^\star l}=\theta_{c^\star l}$. By construction, $\lambda_\epsilon\in B_{c^\star k}(\theta)$, and 
$$
\lim_{\epsilon\to 0} \sum_{l \neq k^{\star}} \alpha_{c^\star l}I^{(c^\star,l)}(\theta,\lambda_\epsilon) = \sum_{l=k^\star+1}^k \alpha_{c^\star l}I(\theta_l,{\mu^\star\over r_l}).
$$
\end{itemize}

From (\ref{eq:inf}), we deduce that constraints (\ref{eq:con1}) are equivalent to (\ref{eq:con2}) (indeed, only the constraints related to $k\in M$ are really active, and for $k\in M$, (\ref{eq:con1}) is equivalent to $\alpha_{c^\star k}I(\theta_{c^\star k},{\mu^\star\over r_k}) \ge 1$). 

\medskip
\noindent
\underline{Constraint (\ref{eq:c2}).} Note that if $k_c^\star< k_0$, then $B_{ck_c^\star}(\theta)=\emptyset$. Assume that $k_c^\star\ge k_0$. When $\lambda\in B_{c k_c^\star}(\theta)$, the optimal (channel, rate) pair under $\lambda$ is $(c, k_c^\star)$. This implies that $r_{k_c^\star}\lambda_{ck_c^\star}\ge \mu^\star$, and so:
$$
\sum_{(c',l) \neq (c^\star,k^{\star})} \alpha_{c'l}I^{(c',l)}(\theta,\lambda)\ge \alpha_{c k_c^\star}I(\theta_{ck_c^\star},{\mu^\star\over r_{k_c^\star}}).
$$
Now select $\lambda_\epsilon$ as follows: $\lambda_{c'k}=\theta_{c'k}$ for all $(c',k)\neq (c,k_c^\star)$, and $\lambda_{ck_c^\star}=\mu^\star/r_{k_c^\star}+\epsilon$. Then $\lambda_\epsilon\in B_{ck_c^\star}(\theta)$ for all $\epsilon>0$, and
$$
\lim_{\epsilon\to 0} \sum_{(c',l) \neq (c^\star,k^{\star})} \alpha_{c'l}I^{(c',l)}(\theta,\lambda)= \alpha_{c k_c^\star}I(\theta_{ck_c^\star},{\mu^\star\over r_{k_c^\star}}).
$$
We conclude that (\ref{eq:c2}) is equivalent to:
$$
\forall c\neq c^\star, \alpha_{ck_c^\star}I(\theta_{ck_c^\star},{  \mu^\star \over r_{k_c^\star}})\ge 1_{k_c^\star\ge k_0}.
$$

\medskip
\noindent
\underline{Constraint (\ref{eq:c3}).} For $k< k_0$, $B_{ck}(\theta)=\emptyset$. Assume that $k\ge k_0$ and $k\neq k_c^\star$. Then $\sum_{(c',l) \neq (c^\star,k^{\star})} \alpha_{c'l}I^{(c',l)}(\theta,\lambda)$ is minimized over $B_{ck}(\theta)$ when for all $c'\neq c$ and all $k$, $\lambda_{c'k}=\theta_{c'k}$, and is actually equal to: $\inf_{\lambda_c\in C_k}\sum_{l}\alpha_{cl}I(\theta_{cl},\lambda_{cl})$. Unfortunately, the above optimization problem cannot be further reduced.
\ep

\section*{Proof of Theorem \ref{th:lower_dep_scaling} }

For $\lambda$: $\forall c$, $\theta_c \in {\cal T}\cap{\cal U}$ and $\alpha \in \RR_+^{C \times K}$, we define:
\als{
	D(\theta,\lambda,\alpha) = \sum_{ (c,k) } \alpha_{ck} I( \theta_{ck},\lambda_{ck}).
}
As defined previously, the ``bad parameter set'' is:
\eqs{
B(\theta) = \{ \lambda \in {\cal T}\cap{\cal U} :  \lambda_{c^\star k^\star} = \theta_{c^\star k^\star}   \;,\; \max_{(c,k)} r_k \lambda_{ck} > \mu^\star   \}.
}
Further define the set:
\eqs{
{\cal C} = \{ \alpha \in \RR_+^{C \times K} : \inf_{\lambda \in B(\theta)} D(\theta,\lambda,\alpha) \geq 1\}.
}

$c_U(\theta)$ in Theorem~\ref{th:lower_dep} is the solution to a minimization problem over ${\cal C}$ . An upper bound of $c_U(\theta)$ is obtained by choosing $\alpha \in {\cal C}$ and by computing the value of the objective function at $\alpha$ which is $\sum_{c,k} \alpha_{c k} ( \mu^\star - \mu_{ck} )$. We prove that if we define $\alpha$ as:
\BIT
\item  $\alpha_{c^\star k} = 1/I( \theta_{c^\star k} , {\mu^\star\over r_k} )$  if  $k \in M$,
\item $\alpha_{c k^\star_c}  = ( \min \{ I( \theta_{c k^\star_c},  {\mu^\star \over r_{ k^\star_c}} ) ,  I( \theta_{c k^\star_c} ,  \theta_{c k^\star_c} - {\delta_c\over r_{ k^\star_c}}) \}    )^{-1}$ if $c \neq c^\star$,
\item  $\alpha_{ck}  = 1/I( \theta_{ck}, \theta_{ck} + {\delta_c \over r_k}  )$  if $c \neq c^\star$ and $k \in M_c$,   	      
\item  $\alpha_{ck} = 0$ if  $c \neq c^\star$ and $k \notin M_c \cup \{ k_c^\star\}$. 	 
\EIT
then $\alpha \in {\cal C}$. To do so, we use the following decomposition: 
$B(\theta) = \cup_{ (c,k) \neq (c^\star , k^\star)} B_{ck}(\theta)$ where
$$
B_{ck}(\theta) = \{ \lambda \in B(\theta) : (c,k) \in \arg \max_{(c' , k')} r_{k'} \lambda_{c' k'} \}.
$$
\medskip
\noindent
\underline{(i) If $\lambda \in \cup_{k \neq k^\star} B_{c^\star k}(\theta)$.} Since $\lambda \in \cup_{k \neq k^\star} B_{c^\star k}(\theta)$, we have $\theta_{c^\star k^\star} = \lambda_{c^\star k^\star }$. Since $k \mapsto r_k \lambda_{c^\star k}$ is unimodal, and $k^\star \notin \arg \max_{k} r_k \lambda_{c^\star k}$, then there must exist a neighbour $k'$ of $k^\star$ such that $ r_{k'} \lambda_{c^\star k'} \ge r_{k^\star} \lambda_{c^\star k^\star} = r_{k^\star} \theta_{c^\star k^\star} = \mu^*$. Hence $\lambda_{c^\star k'} \geq \mu^* / r_{k'}$. Using the monotonicity of the KL divergence: 
\als{
D(\theta,\lambda,\alpha)  &\geq \alpha_{c^\star k'} I(\theta_{c^\star k'} , \lambda_{c^\star k'}  ) \sk
& \geq \alpha_{c^\star k'} I(\theta_{c^\star k'} , \mu^* / r_{k'}  ) \sk
& \geq 1.
}
\medskip
\noindent
\underline{ (ii) If $\lambda \in \cup_{k} B_{c k}(\theta)$ , $c \neq c^\star$.} Under parameter $\lambda$, let $\tilde{k} = \arg \max_{k} r_k \lambda_{c k}$ be the optimal rate for channel $c$. We further consider two cases depending on whether $\tilde{k}$ is equal to $k^\star_c$.\\
Case (a): $\tilde{k} = k^\star_c$. Then $\lambda \in B_{c k^\star_c}(\theta)$, and we have $r_{k^\star_c} \lambda_{c k^\star_c} \geq \mu^\star$. Hence:
\als{
D(\theta,\lambda,\alpha)  & \geq \alpha_{c k^\star_c} I(\theta_{c k^\star_c} , \lambda_{c k^\star_c}  ) \sk
& \geq \alpha_{c k^\star_c} I(\theta_{c k^\star_c} , \mu^\star/r_{k^\star_c} ) \sk
& \geq 1.
}

\noindent
Case (b): $\tilde{k} \neq k^\star_c$. Since $k \mapsto r_k \lambda_{ck}$  is unimodal and $k^\star_c \neq \arg \max_{k} r_k \lambda_{ck}$, there must exist a neigbour $k'$ of $k^\star_c$ such that $ r_{k'} \lambda_{c k'} \ge r_{k^\star_c} \lambda_{c k^\star_c}$. Since $k \mapsto r_k \theta_{ck}$  is unimodal and $k^\star_c = \arg \max_k r_k \theta_{ck}$, we have $r_{k^\star_c} \theta_{c k^\star_c} \geq  r_{k'} \theta_{c k'}$. Therefore:
\als{
 \max(   r_{k^\star_c} | \lambda_{c k^\star_c}  & - \theta_{c k^\star_c} | , r_{k'} | \lambda_{c k'}  - \theta_{c k'} |  ) \\
& \geq  ( r_{k^\star_c} \theta_{c k^\star_c} - r_{k'} \theta_{c k'} )/2 \ge \delta_c.
}
To establish the above inequality, we have used the fact that for all $a , b > 0$ and for all $x \in \RR$, $\max(  |x|  , | x + a + b |  ) \geq (a+b)/2$, and have applied this result for $x=r_{k^\star_c} (\lambda_{c k^\star_c} - \theta_{c k^\star_c})$, $a=r_{k'}\lambda_{ck'}-r_{k_c^\star}\lambda_{ck_c^\star}$, and $b=r_{k_c^\star}\theta_{ck_c^\star}-r_{k'}\theta_{ck'}$. We have shown that:
\begin{quote}
either (b1) $r_{k_c^\star}|\lambda_{ck_c^\star}-\theta_{ck_c^\star}|\ge \delta_c$;\\
or (b2) $r_{k'} | \lambda_{c k'}  - \theta_{c k'} |\ge \delta_c$.
\end{quote}
If (b1) holds, then either $\lambda_{ c k^\star_c} \leq  \theta_{c k^\star_c} - \delta_c/r_{k^\star_c}$ or $\lambda_{ c k^\star_c} \ge  \theta_{c k^\star_c} + \delta_c/r_{k^\star_c}$. In the latter case, we have:
\begin{align*}
\lambda_{ck'}\ge {r_{k_c^\star}\over r_{k'}}\lambda_{ck_c^\star} & \ge {r_{k_c^\star}\over r_{k'}}\theta_{ck_c^\star}+{\delta_c\over r_{k'}}\\
& \ge \theta_{ck'}+{\delta_c\over r_{k'}}.
\end{align*}
If (b2) holds, then either $\lambda_{ c k'} \geq  \theta_{c k'} + \delta_c/r_{k'}$, or $\lambda_{ c k'} \leq  \theta_{c k'} - \delta_c/r_{k'}$. In the latter case, we have:
\begin{align*}
\lambda_{ck_c^\star}\le {r_{k'}\over r_{k_c^\star}}\lambda_{ck'} & \le {r_{k'}\over r_{k_c^\star}}\theta_{ck'}-{\delta_c\over r_{k_c^\star}}\\
&\le \theta_{ck_c^\star} - {\delta_c\over r_{k_c^\star}}.
\end{align*}
In both cases (b1) and (b2), we have proved that either $\lambda_{ c k^\star_c} \leq  \theta_{c k^\star_c} - \delta_c/r_{k^\star_c} $ or $\lambda_{ c k'} \geq  \theta_{c k'} + \delta_c/r_{k'}$. Finally:
\als{
D(\theta,\lambda,\alpha)  & \geq \alpha_{c k^\star_c} I(\theta_{c k^\star_c} , \lambda_{ c k^\star_c} )  +  \alpha_{c k'} I(\theta_{c k'} , \lambda_{ c k'}  ) \sk
&	\geq \max \{  \alpha_{c k^\star_c} I(\theta_{c k^\star_c} , \theta_{c k^\star_c} - {\delta_c\over r_{k^\star_c}} )  ,\\
&\quad\quad\quad   \alpha_{c k'} I(\theta_{c k'} , \theta_{c k'} + {\delta_c\over r_{k'} }) \} \sk   
& \geq 1.
}

We have proved that $\inf_{\lambda\in B(\theta)} D(\theta,\lambda,\alpha)\ge 1$, and thus $\alpha\in {\cal C}$. Now for our choice of $\alpha$, the value of the objective function of the optimization problem in Theorem \ref{th:lower_dep} is $c_u'(\theta)$. We conclude that $c_U(\theta)\le c_U'(\theta)$. \ep

\section*{Proof of Theorem \ref{th:crs}}

We first decompose the regret as the sum of the regret due to exploring the neighbourhood of the optimal rate on each channel, and the other (channel,rate) pairs. By definition of the regret:
\als{
R^{\pi}(T) &= \sum_{(c,k)} (\mu^\star - \mu_{ck}) \EE[ t_{ck}(T) ] \sk
	&= \sum_{c} \sum_{|k-k_c^\star| \leq 1} (\mu^\star - \mu_{ck}) \EE[ t_{ck}(T) ] \sk
	&+ \sum_{c} \sum_{|k-k_c^\star| > 1} (\mu^\star - \mu_{ck}) \EE[ t_{ck}(T) ].
}
Therefore, to prove the theorem, it is sufficient to prove that for all channels $c$ and all $\epsilon >0$:
\als{
 \EE[ t_{ck}(T) ] \leq (1+\epsilon) \log(T) \tau_c^{-1}  +    O(  \log(\log(T)) + \Delta^{-2}),
}
if $|k-k_c^\star| \leq 1$ and:
\eqs{
 \EE[ t_{ck}(T) ] \leq  O(  \log(\log(T)) + \Delta^{-2}).
}
otherwise. To ease notation we define $f(n) = \log(n) + 3 \log(\log(n))$, and the empirical success probability of (channel,rate) pair $(c,k)$ as $\hat\theta_{ck}(n) = \hat\mu_{ck}(n)/\mu_{ck}$.

 Define $A$ the set of instants where there exists at least a (channel, rate) pair $(c,k)$ such that either its index (i.e its upper confidence bound) $q_{ck}(n)$ under-estimates its expected value $\mu_{ck}$, or its lower-confidence bound $\underline{q}_{ck}(n)$ over-estimates its expected value $\mu_{ck}$:
\als{
A^{1} &= \cup_{(c',k')} \{ 1 \leq n \leq T: q_{c'k'}(n) < \mu_{c'k'} \}, \sk
A^{2} &= \cup_{(c',k')} \{ 1 \leq n \leq T: \underline{q}_{c'k'}(n) > \mu_{c'k'} \},\sk
A &= A^{1} \cup A^{2}.
}
 Consider $(c,k)$ and $\epsilon > 0$ both fixed and define $\tilde{t}_{ck} = (1+\epsilon) f(T) \tau_c^{-1}$ if $|k-k_c^\star| \leq 1$ and $\tilde{t}_{ck} = 0$ otherwise. Further define the sets of instants:
\als{
B_{ck} &=  \{ 1 \leq n \leq T: (c,k)(n) = (c,k) , t_{ck}(n) < \tilde{t}_{ck} \} \sk
C_{ck} &=  \{ 1 \leq n \leq T: n \notin A, (c,k)(n) = (c,k) , \sk 
   &  \hspace{4cm} l_c(n) \neq k_{c}^\star \} \sk
D_{ck} &=  \{ 1 \leq n \leq T: n \notin A, (c,k)(n) = (c,k), \sk
   &   \hspace{2cm} l_c(n) = k_{c}^\star, t_{ck}(n) \geq \tilde{t}_{ck} \}.
}
and we have that:
\eqs{
\EE[t_{ck}(T)] \leq \EE[|A|] + \EE[|B_{ck}|]+ \EE[|C_{ck}|]+ \EE[|D_{ck}|]. 
}
At all instants $n \in B_{ck}$, $t_{ck}(n)$ is incremented, therefore $|B_{ck}| \leq \tilde{t}_{ck}$. From the above inequality we deduce:
\eqs{
\EE[t_{ck}(T)] \leq \EE[|A|] + \tilde{t}_{ck}+ \EE[|C_{ck}|]+ \EE[|D_{ck}|].
}
To prove the announced result, it is sufficient to show that the expected size of $A$, $B_{ck}$ and $D_{ck}$ are at most of order $O(\Delta^{-2} + \log(\log(T)))$.

We will prove the following upper bounds:
\begin{itemize}
\item[(i)] $\EE[|A|] = O(\log(\log(T)))$
\item[(ii)] $\EE[|C_{ck}|] = O(\Delta^{-2})$
\item[(iii)] $\EE[|D_{ck}|] = O(1)$.
\end{itemize}

\fbox{ (i) $\EE[|A|] = O(\log(\log(T)))$ }

To upper bound the expected size of $A$, let us prove that: 
\eq{\label{eq:sizeA}
A \subset   \cup_{(c,k)} \{  1 \leq n \leq T:  t_{ck}(n) I(\hat\theta_{ck}(n),\theta_{ck}) \geq f(n) \}.
}
We recall the definition of index  $q_{ck}(n)$:
\eqs{
q_{ck}(n) = \sup \{ q \in [0,r_k]  , t_{ck}(n) I( \hat\theta_{ck}(n) , q/r_k) \leq f(n) \}.
}
Since $q \mapsto I(p,q)$ is increasing for $q \geq p$, if $q_{ck}(n) < \mu_{ck}$ then:
\eqs{
	t_{ck}(n) I(\hat\theta_{ck}(n),\mu_{ck}/r_k) = t_{ck}(n) I(\hat\theta_{ck}(n),\theta_{ck}) \geq f(n).
}
Using the same reasoning for the lower confidence bound $\underline{q}_{ck}(n)$:
\eqs{
\underline{q}_{ck}(n) = \inf \{ q \in [0,r_k]  , t_{ck}(n) I( \hat\theta_{ck}(n) , q/r_k) \leq f(n) \},
}
and since $q \mapsto I(p,q)$ is decreasing for $q \leq p$, if $\underline{q}_{ck}(n) > \mu_{ck}$ then $t_{ck}(n) I(\hat\theta_{ck}(n),\theta_{ck}) \geq f(n)$, so that equation~\eqref{eq:sizeA} is valid.

Applying \cite{garivier2011}[Theorem 10] we have that:
 \eqs{
 \PP[ t_{ck}(n) I(\hat\theta_{ck}(n),\theta_{ck}) \geq f(n) ] \leq e  \frac{ \ceil{ \log(n)f(n)}}{n (\log(n))^{3}}.
 } 

Using a union bound we obtain the announced inequality: 
\als{
\EE[|A|] &\leq  \sum_{n=1}^{T} \sum_{(c,k)} \PP[ t_{ck}(n) I(\hat\theta_{ck}(n),\theta_{ck}) \geq f(n) ] \sk
					&\leq CK e \sum_{n=1}^{T}  \frac{ \ceil{ \log(n)f(n)}}{n (\log(n))^{3}} \sk
					&\leq O(\log(\log(T))).
}

\fbox{ (ii) $\EE[|C_{ck}|] = O(\Delta^{-2})$}

Let us decompose $C_{ck}$ depending on the index of the leader: 
\als{
C_{ck} &= \cup_{k': |k-k'| \leq 1, k' \neq k_{c}^\star} C_{ckk'}, \sk
C_{ckk'} &=  \{ 1 \leq n \leq T: n \notin A, (c,k)(n) = (c,k) , \sk
 & \hspace{3cm} l_c(n) = k' \},
}
the set of instants $n \notin A$ where $(c,k)$ is selected and $k'$ is the leader on channel $c$. 

Fix $k' \neq k_c^{\star}$ and consider $n \in C_{ckk'}$. There exists $\tilde{k}$ such that $|\tilde{k} - k'| = 1$ and $\mu_{c\tilde{k}} \geq \mu_{ck'} + \Delta$ since $k' \neq k_c^\star$ and $k \mapsto \mu_{ck}$ is unimodal for all $c$. Let us prove that we must have $U_c(n) = 0$. Assume that $U_c(n) = 1$ so that $\underline{q}_{ck'}(n) \geq \max_{ k'':|k'' - k'|=1} q_{ck''}(n) \geq q_{c\tilde{k}}(n)$. Since $n \notin A$ we have: $ q_{c\tilde{k}}(n) \geq \mu_{\tilde{k}} > \mu_{k'} \geq  \underline{q}_{ck'}(n)$, a contradiction.

  Define $s= \sum_{n'=1}^{n} \indic \{ n' \in C_{ckk'} \}$ the number of instants in $C_{ckk'}$ between $1$ and $n$. Since for all $n \in C_{ckk'}$ we have $U_c(n) = 0$, we must have $(c,k)(n) \in \arg \min_{ k'':|k'' - k'| \leq 1  } t_{ck''}(n)$ so that $t_{ck''}(n) \geq s/3$ for all $k''$ such that $|k''-k'| \leq 1$. Since  $\mu_{\tilde{k}} \geq \mu_{k'} + \Delta$ and $\hat\mu_{k'}(n) \geq \hat\mu_{\tilde{k}}(n)$ , we must have either  $| \hat\mu_{k'}(n) - \mu_{k'} | \geq \Delta/2  $ or $| \hat\mu_{\tilde{k}}(n) - \mu_{\tilde{k}} | \geq \Delta/2$. 
  
   So $t_{ck''}(n) \geq s/3$ for all $k''$ such that $|k' - k''| \leq 1$ and  $ \max_{k'': |k' - k''| \leq 1 } | \hat\mu_{k''}(n) - \mu_{k''}(n) | \geq \Delta/2$, and applying Lemma~\ref{lem:deviation}, we obtain that $\EE[|C|] = O(\Delta^{-2})$.
  
  \fbox{ (iii) $\EE[|D_{ck}|] = O(1)$ }
  
  First it is noted that if $|k - k_c^\star| > 1$, then $D_{ck} = \emptyset$, since $n \in D$ implies $l_c(n) = k_c^\star$ and by design of CRS-T: $|k(n) - k_c^\star| = |k(n) - l_{c(n)}(n)| \leq 1$.
  
 Now consider $k$ such that $|k -k_c^\star| \leq 1$ and decompose $D_{ck}$ depending on the value of the test $U_c(n) \in \{0,1\}$:
 \als{
  D_{ck} &= D_{ck}^{0} \cup D_{ck}^{1} , \sk
  D_{ck}^{0} &= \{  1 \leq n \leq T: n \in D_{ck}, U_c(n) = 0   \} , \sk
  D_{ck}^{1} &= \{ 1 \leq n \leq T: n \in D_{ck}, U_c(n) = 1   \}.
 }
 Consider $n \in D_{ck}^{1}$. Since $l_c(n) = k^\star_c$ and $U_c(n) = 1$, by design of CRS-T we have $U_{c'}(n) = 1$ for all channels $c'$ otherwise $c$ is not selected. By the same reasoning as above, since $n \notin A$, $U_{c'}(n) = 1$ for all $c'$ implies $l_{c'}(n) = k_{c'}^\star$ for all $c'$. Also, by design of CRS-T $k = k^\star_c$ otherwise $(c,k)$ is not selected. Since $(c,k)$ is selected  $q_{ck_c^\star}(n) \geq   q_{c^\star l_{c^\star}(n)}(n) = q_{c^\star k^\star}(n) \geq \mu^\star$ since  $n \notin A$. By definition of the index $q$, $q_{ck_c^\star}(n) \geq \mu^\star$ implies:
 \eqs{
 	t_k(n) I( \hat\theta_{ck}(n) , \mu^\star/r_k )  \leq f(n) \leq f(T)
 }
 since $q \mapsto I(p,q)$ is increasing for $q \geq p$ and $n \mapsto f(n)$ is increasing. By definition of $\tau_c$ we have:
\als{
  t_k(n) &\geq \tilde{t}_{ck} \sk
  &= (1+\epsilon)f(T) \tau_c^{-1} \sk 
  &\geq (1+\epsilon)f(T) / I(\mu_{ck},\mu^\star/r_k),
  } 
  so that:
  \eqs{
  	 I( \hat\theta_{ck}(n) , \mu^\star/r_k )  \leq  I(\theta_{ck},\mu^\star/r_k)/(1+\epsilon).
  } 
  Therefore, since $p \mapsto I(p,q)$ is decreasing for $p \leq q$ there exists $\eta > 0$ (depending on $\epsilon$) such that:
  \eqs{
   |\hat\theta_{ck}(n) - \theta_{ck}| \geq \eta.
  }
  Therefore:
  \eqs{
  	D_{ck}^{1} \subset \{  (c,k)(n) = (c,k) ,   |\hat\theta_{ck}(n) - \theta_{ck}| \geq \eta \}
  }
  
 so that $\EE[D_{ck}^{1}] = O(\eta^{-2})$ applying Lemma~\ref{lem:deviation} once again.
 
 We turn to $D_{ck}^{0}$. Since $U_c(n) = 0$, we have $  \underline{q}_{ck_{c}^\star}(n) \leq \max_{k': |k'-k_c^\star| = 1} q_{ck'}(n)$ so that either (a) $\underline{q}_{ck_{c}^\star}(n) \leq \tilde{\mu}_c$ or (b) $\max_{k': |k'-k_c^\star| = 1} q_{ck'}(n) \geq \tilde{\mu}_c$.
 
 \underline{Case (a)}: if $\underline{q}_{ck_{c}^\star}(n) \leq \tilde{\mu}_c$ we have:
 \eqs{
 	t_{ck_c^{\star}}(n) I( \hat\theta_{c k_c^{\star}}(n), \tilde{\mu}_c/r_{k_c^{\star}}  ) \leq f(n) \leq f(T),
 }
 and since $t_{ck_c^{\star}}(n) \geq (1+\epsilon) f(T) / I( \theta_{c k_c^{\star}}, \tilde{\mu}_c/r_{k_c^{\star}})$: 
 \eq{\label{eq:kl1}
 I( \hat\theta_{c k_c^{\star}}(n) , \tilde{\mu}_c/r_{k_c^{\star}} ) \leq I( \theta_{c k_c^{\star}} , \tilde{\mu}_c/r_{k_c^{\star}} )/(1+\epsilon).
 }
 \underline{Case (b)}: consider $k'$ such that $|k'-k_c^\star| = 1$, if $q_{ck'}(n) \geq \tilde{\mu}_c$, then:
  \eqs{
 	t_{ck'}(n) I( \hat\theta_{c k'}(n), \tilde{\mu}_c/r_{k'}  ) \leq f(n) \leq f(T),
 }
 and since $t_{ck'}(n) \geq (1+\epsilon) f(T) / I( \theta_{c k'}, \tilde{\mu}_c/r_{k'})$:
  \eq{\label{eq:kl2}
 I( \hat\theta_{c k'}(n) , \tilde{\mu}_c/r_{k'} ) \leq I( \theta_{c k'}(n) , \tilde{\mu}_c/r_{k'} )/(1+\epsilon).
 }
Putting \eqref{eq:kl1} and \eqref{eq:kl2} together, we deduce that there must exist $\eta$ such that:
 \eqs{
   \sup_{k': |k'-k_c^\star| \leq 1} |\hat\theta_{ck'}(n) - \theta_{ck'}| \geq \eta,
 }
and therefore:
\als{
  D_{ck}^{0} \subset \{   (c,k)(n) &= \arg \min_{k':|k'- k_c^\star| \leq 1 } t_{ck'}(n), \sk
  									& \sup_{k':|k'- k_c^\star| \leq 1}  |\hat\theta_{ck'}(n) - \theta_{ck'}| \geq \eta \},
}
and using Lemma~\ref{lem:deviation} a third time we have that $\EE[|D^{0}_{ck}|] = O(\eta^{-2})$ which concludes the proof. \ep
  
\section*{A deviation result}

The following result proven in \cite{combesICMLTR2014}[Lemma 2.2] is reproduced here for completeness.

\begin{lemma}[\cite{combesICMLTR2014}]\label{lem:deviation}
Let $\epsilon > 0$. Consider $( X(t) )_{t \geq 0}$ i.i.d. random variables in $[0,1]$ with common expectation $\mu$. Define ${\cal F}_n$ the $\sigma$-algebra generated by $( X(t) )_{1 \leq t \leq n}$. Consider a random variable $B_t \in \{0,1\}$ such that $B_t$ is ${\cal F}_{t-1}$ measurable for all $t \geq 0$, and define $t(n) = \sum_{t=1}^n B_t$ and $\hat\mu(n) = (1/t(n)) \sum_{t=1}^n  B_t X_t$. Let $\Lambda \subset {\mathbb N}$ be a (random) set of instants. Assume that there exists a sequence of (random) sets $(\Lambda(s))_{s\ge 1}$ such that (i) $\Lambda \subset \cup_{s \geq 1} \Lambda(s)$, (ii) for all $s\ge 1$ and all $n\in \Lambda(s)$, $t(n) \ge \epsilon s$, (iii) $|\Lambda(s)| \leq 1$, and (iv) the event $n \in \Lambda(s)$ is ${\cal F}_n$-measurable. Then for all $\delta > 0$:
\eqs{
\EE[ \sum_{n \geq 1} \indic\{ n \in \Lambda , |\hat\mu(n) - \mu| > \delta \}]  \leq  \frac{1}{\epsilon \delta^2}.
}
\end{lemma}


\end{document}